\documentclass[reqno,12pt]{article}

\usepackage{mathtools}
\usepackage{color}
\usepackage{subcaption}

\usepackage{xstring}

\def \nn{\nonumber}

\def \h#1{\widehat{#1}}
\def \t#1{\widetilde{#1}}
\def \xht#1{\widehat{\widetilde{#1}}}
\def \b#1{\overline{#1}}
\def \xbt#1{\overline{\widetilde{#1}}}
\def \xbh#1{\overline{\widehat{#1}}}
\def \xbht#1{\overline{\widehat{\widetilde{#1}}}}

\usepackage{amsmath,amssymb,epic}

\newcommand{\bse}{\begin{subequations}}
\newcommand{\ese}{\end{subequations}}
\newcommand{\oeq}{{\,\stackrel{\circ}{=}\,}}

\newtheorem{definition}{Definition}

\usepackage{ifthen}

\usepackage{tikz}
\usepackage{tikz-3dplot}

\usetikzlibrary{positioning,calc}

\newcommand\xyband[7]{
\draw[cube, very thin, opacity=#7, fill=#6] %
(#2,#3,#1) -- (#2,#5,#1) -- (#4,#5,#1) -- (#4,#3,#1) -- (#2,#3,#1);
}
\newcommand\xzband[7]{
\draw[cube, very thin, opacity=#7, fill=#6] %
(#2,#1,#3) -- (#2,#1,#5) -- (#4,#1,#5) -- (#4,#1,#3) -- (#2,#1,#3);
}
\newcommand\yzband[7]{
\draw[cube, very thin, opacity=#7, fill=#6] %
(#1,#2,#3) -- (#1,#2,#5) -- (#1,#4,#5) -- (#1,#4,#3) -- (#1,#2,#3);
}

\newcommand\crosspq{%
\draw[cube, very thick, brown] (0,0,0) -- (2,2,0);
\draw[cube, very thick, brown
] (2,0,0) -- (0,2,0);
}
\newcommand\crossqr{%
\draw[cube, very thick, brown] (0,0,0) -- (0,2,2);
\draw[cube, very thick, brown] (0,0,2) -- (0,2,0);
}
\newcommand\crossrp{%
\draw[cube, very thick, brown] (0,0,0) -- (2,0,2);
\draw[cube, very thick, brown] (2,0,0) -- (0,0,2);
}

\newcommand\edgespq[2]{%
{\ifthenelse{\equal{#1}{q}}{%
\xyband{#2}{0}{0}{0.3}{2}{red}{0.6};
\xyband{#2}{2}{0}{1.7}{2}{red}{0.6};}{%
\xyband{#2}{0}{0}{2}{0.2}{red}{0.6};
\xyband{#2}{0}{2}{2}{1.8}{red}{0.6};
}}}

\newcommand\edgespr[2]{%
{\ifthenelse{\equal{#1}{r}}{%
\xzband{#2}{0}{0}{0.3}{2}{blue}{0.6};
\xzband{#2}{2}{0}{1.7}{2}{blue}{0.6};}{%
\xzband{#2}{0}{0}{2}{0.2}{blue}{0.6};
\xzband{#2}{0}{2}{2}{1.8}{blue}{0.6};
}}}

\newcommand\edgesqr[2]{%
{\ifthenelse{\equal{#1}{r}}{%
\yzband{#2}{0}{0}{0.2}{2}{green}{0.6};
\yzband{#2}{2}{0}{1.8}{2}{green}{0.6};}{%
\yzband{#2}{0}{0}{2}{0.2}{green}{0.6};
\yzband{#2}{0}{2}{2}{1.8}{green}{0.6};
}}}

\newcommand\laatikko[0]{%
        \draw[cube] (0,0,0) -- (0,2,0) -- (2,2,0) -- (2,0,0) -- cycle;
        \draw[cube] (0,0,2) -- (0,2,2) -- (2,2,2) -- (2,0,2) -- cycle;
        \draw[cube] (0,0,0) -- (0,0,2);
        \draw[cube] (0,2,0) -- (0,2,2);
        \draw[cube] (2,0,0) -- (2,0,2);
        \draw[cube] (2,2,0) -- (2,2,2);%
}
\tikzstyle{block}=[draw,  rectangle, 
    minimum height=1em, minimum width=3em]

\newcommand{\hajota}[1]{\{\StrLeft{#1}{2},\StrMid{#1}{3}{4},\StrMid{#1}{5}{6}\}}
\newcommand{\hajotalong}[1]%
{\{\StrLeft{#1}{2},\StrMid{#1}{3}{4},\StrMid{#1}{5}{6}\}-\StrMid{#1}{8}{8}}

\newcommand{\ynum}[1]%
{$\displaystyle{\genfrac{}{}{0pt}{}{\text{\hajota{#1}}}{\eqref{#1}}}$}

\newcommand{\ynumx}[3]%
{$\displaystyle{\genfrac{}{}{0pt}{}%
{\text{\hajota{#1}}}{\eqref{#1}\ \frac{#2}{#3}}}$}

\newcommand{\ynumxw}[3]%
{$\displaystyle{\genfrac{}{}{0pt}{}%
{\text{\hajota{#1}}}{<\!\!\eqref{631010}\ \frac{#2}{#3}}}$}

\newcommand{\ynumlongx}[3]%
{$\displaystyle{\genfrac{}{}{0pt}{}%
{\text{\hajotalong{#1}}}{\eqref{#1}\ \frac{#2}{#3}}}$}

\newcommand{\redudown}[3]%
{\node [block,below of = #1] (#2) {\ynum{#2}};
 \path[->] (#1) edge node {#3} (#2);}

\newcommand{\reduright}[3]%
{\node [block,right of = #1,node distance=3.5cm] (#2) {\ynum{#2}};
 \path[->] (#1) edge node {#3} (#2);}

\newcommand{\redudownx}[5]%
{\node [block,below of = #1] (#2) {\ynumx{#2}{#4}{#5}};
 \path[->] (#1) edge node {#3} (#2);}

\newcommand{\redurightx}[5]%
{\node [block,right of = #1,node distance=3.5cm] (#2) {\ynumx{#2}{#4}{#5}};
 \path[->] (#1) edge node {#3} (#2);}

\newcommand{\reduleftx}[5]%
{\node [block,left of = #1,node distance=3.5cm] (#2) {\ynumx{#2}{#4}{#5}};
 \path[->] (#1) edge node[above] {#3} (#2);}

\newcommand{\reduleftxw}[5]%
{\node [block,left of = #1,node distance=3.5cm] (#2) {\ynumxw{#2}{#4}{#5}};
 \path[->] (#1) edge node[above] {#3} (#2);}

\newcommand{\redurightlong}[3]%
{\node [block,right of = #1,node distance=3.5cm] (#2) {\ynumlong{#2}};
 \path[->] (#1) edge node {#3} (#2);}

\newcommand{\reduleftlong}[3]%
{\node [block,left of = #1,node distance=3.5cm] (#2) {\ynumlong{#2}};
 \path[->] (#1) edge node[above] {#3} (#2);}

\newcommand{\reduleftlongx}[5]%
{\node [block,left of = #1,node distance=3.5cm] (#2) {\ynumlongx{#2}{#4}{#5}};
 \path[->] (#1) edge node[above] {#3} (#2);}

\newcommand{\redudownlong}[3]%
{\node [block,below of = #1] (#2) {\ynumlong{#2}};
 \path[->] (#1) edge node {#3} (#2);}

\newcommand{\redudownlongx}[5]%
{\node [block,below of = #1] (#2) {\ynumlongx{#2}{#4}{#5}};
 \path[->] (#1) edge node {#3} (#2);}

\title{Search for CAC-integrable homogeneous quadratic\\ triplets of quad
  equations and\\  their classification by BT and Lax} \author{Jarmo
  Hietarinta\footnote{E-mail: jarmo.hietarinta@utu.fi}
  \\Department of Physics and Astronomy\\ University of Turku,
  FIN-20014 Turku, Finland}

\begin{document}

\maketitle

\begin{abstract}
We consider two-dimensional lattice equations defined on an elementary
square of the Cartesian lattice and depending on the variables at the
corners of the quadrilateral. For such equations the property often
associated with integrability is that of ``multidimensional
consistency'' (MDC): it should be possible to extend the equation from
two to higher dimensions so that the embedded two-dimensional lattice
equations are compatible. Usually compatibility is checked using
``Consistency-Around-a-Cube'' (CAC).  In this context it is often
assumed that the equations on the six sides of the cube are the same
(up to lattice parameters), but this assumption was relaxed in the
classification of Boll \cite{Boll2011}. We present here the results of
a search and classification of homogeneous quadratic triplets of
multidimensionally consistent lattice equations, allowing different
equations on the three orthogonal planes (hence triplets) but using
the same equation on parallel planes. No assumptions are made about
symmetry or tetrahedron property. The results are then grouped by
subset/limit properties, and analyzed by the effectiveness of their
B\"acklund transformations, or equivalently, by the quality of their
Lax pair (fake or not).
\end{abstract}

\section{Introduction}
\subsection{General setting}
When we discuss integrable equations, continuous or discrete, there is
always the question of which definition of integrability should be
used. Unfortunately at this time it is still not possible to give a
universal definition of integrability. This is because although
integrability can be discussed for various different classes of
equations, it manifests itself in different form depending on the
class. In place of an elusive universal definition, we do have many
different specific properties that are associated with equations that
are considered integrable. These properties depend on the class of
equations.\footnote{ For example the existence of $N$-soliton
  solutions for $(1+1)$ dimensional PDE's is strongly associated with
  integrability, but it does not even make sense for Hamiltonian
  mechanical systems.} However, there are some properties that can be
applied to a variety of classes of equations, such as behavior around
singularities, existence of a Lax pair with spectral parameter, or the
existence of sufficient number of conservation laws, and these have
sometimes been elevated as definitions of integrability, while some
other proposed definitions can only be taken as indicators, or
necessary conditions.

In this paper we consider difference equations defined on an elementary square of
the Cartesian lattice, with the dynamical variables located on the
corners of the square, so called quadrilateral equations. The
following basic assumptions are made:
\begin{definition}[Acceptable equations]:\label{D:1}
\begin{enumerate}
\item The equation depends on all corner variables of the quadrilateral.
\item The equation is affine linear in each corner variable.
\item The equation is irreducible.
\end{enumerate}
\end{definition}
In this paper we assume furthermore that the equations are
{\em homogeneous quadratic}.

Within the present context of quadrilateral equations there are again
several properties that are strongly associated with
integrability. One is obtained using algebraic entropy
\cite{Viallet2006}, that is, the growth of complexity under evolution:
If growth is linear the equation is linearizable, if the growth is
polynomial the equation is integrable, if the growth is exponential
the equation shows chaotic behavior. Other information is obtained
from symmetry analysis \cite{LW06symmrev}. 

The properties just mentioned involve analysis on the 2D
lattice, but there is also a criterion that is based on a
multidimensional lattice. It is well known that soliton equations come
in hierarchies, with infinite number of different evolutionary times,
and in the context of lattice equations this has been associated with
multidimensionality \cite{NRGO01}. 

In such an approach the original 2D quadrilateral equation is extended
into $\mathbb Z^3$ by introducing accompanying compatible equations on
the other 2D-planes of the 3D-space. In the strictest version the
accompanying equations are obtained from the original just by changing
some lattice variables and parameters. This approach was used by
Adler, Bobenko and Suris (together with the additional assumptions of
symmetry and the ``tetrahedron property'' (TET)) in order to obtain a
classification, the ``ABS list'' \cite{ABS03}, which has been
important in resurrecting the study of integrable lattice equations.
A more relaxed version is to allow completely different quad-equations
on the three different $\mathbb Z^2$ lattice planes, this approach was
taken by Boll in his classification \cite{Boll2011}, where the
tetrahedron condition was also used extensively.  Even more generally,
the accompanying equations could also live in bigger stencils (this is
needed if the Lax matrix is bigger that $2\times2$.)

In any case the 3D system of equations must be compatible. If all
equations are quad equations, the compatibility condition is called
``Consistency-Around-a-Cube'' (CAC) and is elaborated below. One
statement relating 2D-integrability and CAC is that if an equation has
been found to be integrable by a 2D condition, then such a compatible
extension should be possible.\footnote{ For example, the $Q_V$
  equation found by Viallet using algebraic entropy analysis
  \cite{Viallet09} has many free parameters and when that equation is
  extended to 3D one is led to the parameterization of Adler's $Q_4$
  equation \cite{Adl98}.}

In the present work we take completely free acceptable
(by Definition \ref{D:1}) quadratic equations on the three $\mathbb
Z^2$ lattices of $\mathbb Z^3$ and classify those that have CAC.

\subsection{Detailed formulation}
\begin{figure}
\centering
\begin{tikzpicture}[scale=1.90,tdplot_main_coords, cube/.style={black}]
  \draw[cube] (0,0,0) -- (0,2,0) -- (2,2,0) -- (2,0,0) -- cycle;
        \draw[cube] (0,0,0) -- (0,0,2);
        \draw[cube] (0,2,0) -- (0,2,2);
        \draw[cube] (2,0,0) -- (2,0,2);
\draw[cube] (0,0,2) -- (0,2,2);
\draw[cube] (0,0,2) -- (2,0,2);
\draw[cube] (0,0,0) -- (2,0,2);
\draw[cube] (0,0,0) -- (0,2,2);
\draw[cube] (0,0,0) -- (2,2,0);
\draw[cube] (0,0,2) -- (2,0,0);
\draw[cube] (0,0,2) -- (0,2,0);
\draw[cube] (2,0,0) -- (0,2,0);
\node at (0,0.1,0.2) {$x$};
\node at (2,0,-0.2) {$\t x$};
\node at (2,2,-0.2) {$\xht x$};
\node at (0,2,-0.2) {$\h x$};
\node at (0,0.2,2.2) {$\b x$};
\node at (2,0,2.2) {$\xbt x$};
\node at (0,2,2.2) {$\xbh x$};
\node at (0,0.15,0.8) {$c_5$};
\node at (0,1.85,1) {$c_6$};
\node at (0,1,1.85) {$c_3$};
\node at (0,1,0.15) {$c_1$};
\node at (0,0.5,1.65) {$c_2$};
\node at (0,1.50,1.65) {$c_4$};
 \fill[white,fill opacity=1] (0,0.8,0.8) rectangle (0,1.2,1.2);
\node at (0,1,1) {$(q,r)$};

\node at (0.95,0.185,0) {$c_1$};
\node at (1.1,1.8,0) {$c_3$};
\node at (0.25,1,0) {$c_5$};
\node at (1.72,1,0) {$c_6$};
\node at (0.42,0.25,0) {$c_4$};
\node at (0.3,1.7,0) {$c_2$};
 \fill[white,fill opacity=1] (0.4,0.45,0) rectangle (1.4,1.45,0);
\node at (0.9,0.95,0) {$(p,q)$};

\node at (0.25,0,1.1) {$c_1$};
\node at (1.75,0,0.9) {$c_3$};
\node at (1,0,0.15) {$c_5$};
\node at (1,0,1.87) {$c_6$};
\node at (1.65,0,1.75) {$c_4$};
\node at (0.45,0,1.55) {$c_2$};
 \fill[white,fill opacity=1] (0.8,0,0.8) rectangle (1.25,0,1.22);
\node at (1,0,1) {$(r,p)$};

\end{tikzpicture}\hspace{0.5cm}\begin{tikzpicture}[scale=1.50,tdplot_main_coords, cube/.style={very thick,black}]
\laatikko
\node at (0,0,-0.2) {$x$};
\node at (2,0,-0.2) {$\t x$};
\node at (2,2,-0.2) {$\xht x$};
\node at (0,2,-0.2) {$\h x$};
\node at (0,0.2,2.2) {$\b x$};
\node at (2,0,2.3) {$\xbt x$};
\node at (2,2,2.3) {$\xbht x$};
\node at (0,2,2.3) {$\xbh x$};
\draw[thick,->] (0,0,0) -- (3,0,0);
\draw[thick,->] (0,0,0) -- (0,2.7,0);
\draw[thick,->] (0,0,0) -- (0,0,2.7);
\node at (0,0.4,2.7) {$r,\, k$};
\node at (3,-0.4,0) {$p,\, n$};
\node at (0,2.7,0.2) {$q,\, m$};
\end{tikzpicture}
\caption{The consistency cube. The picture on the right gives the location
  of the variables, lattice directions and parameters. The picture on
  the left explains the coefficients: the variables of a product are
  connected by a line, adjacent to which is given the name of the
  corresponding coefficient function. The six functions within a
  quadrilateral depend on the lattice parameters given at the center
  of the quadrilateral.  \label{F:CACcube}}
\end{figure}
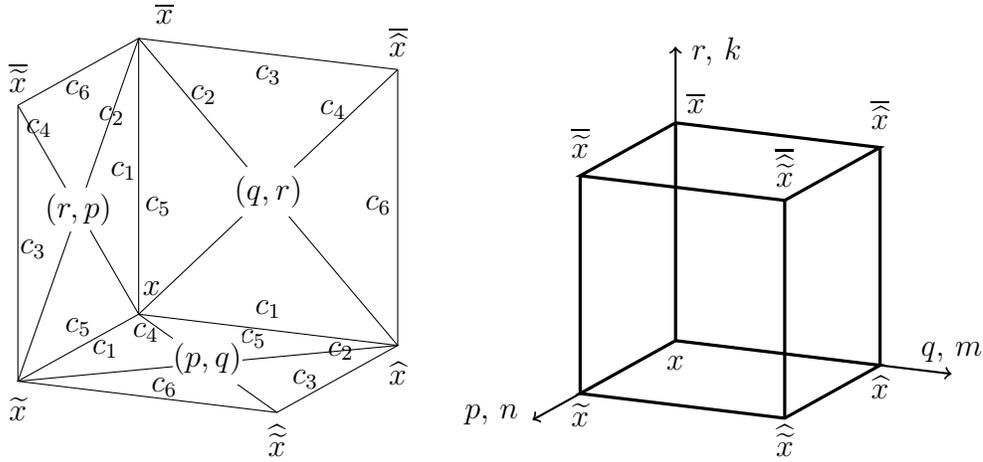

For consistency analysis we consider a 3D cube and assign
equations on each side of the cube. Furthermore we will extend the
consideration from one cube to the full $\mathbb Z^3$ lattice and
embed the equation as follows:
\begin{enumerate}
\item On a given $\mathbb Z^2$ plane all elementary squares have
  equations of the same form (that is, only the corner variables
  change corresponding to the location). The coefficients of the
  equation may depend on the two lattice parameters associated with
  the plane but not on the location.
\item When extended to the $\mathbb Z^3$ lattice, the quadrilaterals
  on parallel planes all carry the same equation but intersecting planes
  may have different equations.
\end{enumerate}
In $\mathbb Z^2$ the corner variables are usually notated as $x_{n,m}$
where $(n,m)$ gives the locations on the Cartesian plane, while in
$\mathbb Z^3$ the variables are indexed as $x_{n,m,k}$. When dealing
with a specific quadrilateral or cube we use shorthand notation
\begin{align*}
&x_{n,m,k}=x,\,x_{n+1,m,k}=\t x,\,x_{n,m+1,k}=\h x,\,x_{n,m,k+1}=\b x,\\
&x_{n+1,m+1,k}=\xht x,\,x_{n+1,m,k+1}=\xbt x,\,x_{n,m+1,k+1}=\xbh x,\
x_{n+1,m+1,k+1}=\xbht x.
\end{align*}
An acceptable quadratic quadrilateral equation can then be written as
\begin{equation}\label{eq:perus}
Q_{12}(x,\t x,\h x,\xht x;p,q):= x\t x\,c_1+\t x \h x c_2+\h x\xht
x\,c_3+ x\xht x\, c_4+x \h x c_5+ \t x\xht x c_6=0,
\end{equation}
where the $c_j$ may depend on the lattice variables $p,q$, which are
associated with the $n,m$ or tilde and hat directions, respectively. The
lattice variable for $k$ or bar direction is $r$.  For the consistency
cube we need $ Q_{12}(x,\t x,\h x,\xht x;p,q)=0$, $ Q_{23}(x,\h x,\b
x,\xbh x;q,r)=0$, and $ Q_{31}(x,\b x,\t x,\xbt x;r,p)=0$, that is
\bse\begin{eqnarray} x\t x\,c_1(p,q)+\t x \h x
c_2(p,q)+\h x\xht x\,c_3(p,q) + x\xht x\, c_4(p,q)+x \h x c_5(p,q)+ \t
x\xht x\,c_6(p,q)&=&0,\label{eq:botx}\\
 x\h x\,c_1(q,r)+\h x \b x c_2(q,r)+\b x\xbh
x\,c_3(q,r) + x\xbh x\, c_4(q,r)+x \b x c_5(q,r)+ \h x\xbh x\,
c_6(q,r)&=&0,\label{eq:bacx}\\ x\b x\,c_1(r,p)+\b x \t x c_2(r,p)+\t x\xbt
x\,c_3(r,p) + x\xbt x\, c_4(r,p)+x \t x c_5(r,p)+ \b x\xbt
x\,c_6(r,p)&=&0,\phantom{mm}\label{eq:lefx}
\end{eqnarray}\ese
respectively. We also need their shifts.  When writing this set of
equations we have used cyclic convention $p \to q \to r \to p$ and
$\ \, \t {\phantom x}\, \to \, \h {\phantom x}\, \to\, \b {\phantom
  x}\,\to\, \t {\phantom x}\,$, see also Figure \ref
{F:CACcube}. Note that the names of the lattice variables in $c_i$ are
also used for function names. That means, for example, that there is
no $c_1(x,y)$ from which $c_1(p,q)$ and $c_1(q,r)$ could be obtained,
rather $c_1(p,q)$ and $c_1(q,r)$ are independent functions.

\tdplotsetmaincoords{75}{115}

Under the assumptions above the consistency conditions arise as follows:
The equations on the faces of the consistency cube are given by:
\bse\label{eq:conseqs}\begin{align}
\text{bottom:}\quad & Q_{12}(x,\t x,\h x,\xht x;p,q)=0.&
\text{top:}\quad &Q_{12}(\b x,\xbt x,\xbh x,\xbht x;p,q)=0,&\\ 
\text{back:}\quad & Q_{23}(x,\h x,\b x,\xbh x;q,r)=0,&
\text{front:} \quad &Q_{23}(\t x,\xht x,\xbt x,\xbht x;q,r)=0,&\\
\text{left:}\quad & Q_{31}(x,\b x,\t x,\xbt x;r,p)=0,&
\text{right:} \quad &Q_{31}(\h x,\xbh x,\xht x,\xbht x;r,p)=0.&
\end{align}
\ese We need precisely four initial values to compute the other
values, three of the initial values should be on one quadrilateral and
the fourth somewhere else.  The standard way (there are others) is to
take $x,\, \t x,\, \h x,\,\b x$ as the initial values and solve the
LHS equations of \eqref{eq:conseqs} for $\xht x,\, \xbh x$ and $\xbt
x$ and substitute them to the RHS. Then from each of the three RHS
equations we can solve for $\xbht x$ and the results should be the
same, therefore we get two conditions for consistency.

\subsection{Organization of the search\label{S:org}}
The condition that the different ways to compute $\xbht x$ provide the
same result leads to a set of equations which are polynomials in the
initial values $x,\, \t x,\, \h x,\,\b x$. The coefficient of every
monomial $x^\alpha\, \t x^\beta\, \h x^\gamma\,\b x^\delta$ has to
vanish separately and this yields polynomial equations on
$c_i(p,q),\,c_i(q,r),c_i(r,p),\,\forall i=1,\dots, 6$.

In analyzing the ensuing equations the solution process branches
depending on whether some coefficient function is zero or not. It is
not easy to keep track of such branching and therefore we choose here
a different approach.

We will do a pre-analysis by setting some coefficients in
\eqref{eq:perus} to zero. For that purpose let us write
\eqref{eq:perus} as
\begin{equation}\label{eq:perus-io}
\iota_1 x\t x\,c_1+\iota_2 \t x \h x c_2+\iota_3 \h x\xht x\,c_3+
\iota_4 x\xht x\, c_4+\iota_5 x \h x c_5+ \iota_6 \t x\xht x c_6=0,
\end{equation}
where $\iota_j\in\{0,1\}$ while $c_j$ are all nonzero. By various
choices of $\iota_j$ we can have $2^6=64$ equations, but by the
conditions in Definition \ref{D:1} this actually yields 37
equations. In order to keep track of the different cases it is
convenient to name the equations by the list
$\{\iota_1,\iota_2,\iota_3,\iota_4,\iota_5,\iota_6\}$ and if we
consider this as a binary number then we can construct also the
corresponding decimal number:
$N=\iota_1+2\iota_2+4\iota_3+8\iota_4+16\iota_5+32\iota_6$. Then we
have for example
\[
 x\t x c_1+\t x \h x c_2+x\xht x\, c_4\quad \leadsto\quad \{1,1,0,1,0,0\}=11.
\]
With this coding the list of the 37 equations passing the conditions
1-3 of Definition \ref{D:1} is given by
\begin{align*}
&\{\phantom{2}  5,\{1,0,1,0,0,0\}\}, \{\phantom{2} 7,\{1,1,1,0,0,0\}\},
\{10,\{0,1,0,1,0,0\}\}, \{11,\{1,1,0,1,0,0\}\},\\&
\{13,\{1,0,1,1,0,0\}\}, \{14,\{0,1,1,1,0,0\}\},
\{15,\{1,1,1,1,0,0\}\}, \{21,\{1,0,1,0,1,0\}\},\\&
\{23,\{1,1,1,0,1,0\}\}, \{26,\{0,1,0,1,1,0\}\},
\{27,\{1,1,0,1,1,0\}\}, \{29,\{1,0,1,1,1,0\}\},\\&
\{30,\{0,1,1,1,1,0\}\}, \{31,\{1,1,1,1,1,0\}\},
\{37,\{1,0,1,0,0,1\}\}, \{39,\{1,1,1,0,0,1\}\},\\&
\{42,\{0,1,0,1,0,1\}\}, \{43,\{1,1,0,1,0,1\}\},
\{45,\{1,0,1,1,0,1\}\}, \{46,\{0,1,1,1,0,1\}\},\\&
\{47,\{1,1,1,1,0,1\}\}, \{48,\{0,0,0,0,1,1\}\},
\{49,\{1,0,0,0,1,1\}\}, \{50,\{0,1,0,0,1,1\}\},\\&
\{51,\{1,1,0,0,1,1\}\}, \{52,\{0,0,1,0,1,1\}\},
\{53,\{1,0,1,0,1,1\}\}, \{54,\{0,1,1,0,1,1\}\},\\&
\{55,\{1,1,1,0,1,1\}\}, \{56,\{0,0,0,1,1,1\}\},
\{57,\{1,0,0,1,1,1\}\}, \{58,\{0,1,0,1,1,1\}\},\\&
\{59,\{1,1,0,1,1,1\}\}, \{60,\{0,0,1,1,1,1\}\},
\{61,\{1,0,1,1,1,1\}\}, \{62,\{0,1,1,1,1,1\}\},\\& \{63,\{1,1,1,1,1,1\}\}
\end{align*}
Note that 5, 10 and 48 yield two-term equations.

In order to populate the consistency cube we need triplets of
equations, leading in principle to $37^3=50653$ cases, fortunately
this number can be reduced by applying symmetries.

\section{Symmetries}
For triplets of equations we use a list $\{X,Y,Z\}$ where $X$ is the
decimal code for the bottom equation, $Y$ for the back equation and
$Z$ for the left equation. The top, front and right equations are
obtained by a coordinate shift and therefore there is no need to
enumerate them separately.

We will now look how some basic symmetries that do not change the CAC
property operate on the codes. For easy reference we write here the
triplet with $\iota$ but without lattice parameters:
\bse\label{eq:seqs}\begin{eqnarray} \iota_1 x\t x\,c_1+\iota_2\t x \h
  x c_2 +\iota_3\h x\xht x\,c_3 + \iota_4x\xht x\, c_4 +\iota_5x \h x
  c_5 +\iota_6 \t x\xht x\,c_6 &=&0,\label{eq:sbotx}\\ \iota_1 x\h
  x\,c_1 +\iota_2\h x \b x c_2 +\iota_3\b x\xbh x\,c_3 +\iota_4 x\xbh
  x\, c_4 +\iota_5x \b x c_5 + \iota_6\h x\xbh x\, c_6
  &=&0,\label{eq:sbacx}\\ \iota_1x\b x\,c_1 +\iota_2\b x \t x c_2
  +\iota_3\t x\xbt x\,c_3 + \iota_4x\xbt x\, c_4 +\iota_5x \t x c_5 +
  \iota_6 \b x \xbt x\,c_6&=&0.\label{eq:slefx}
\end{eqnarray}\ese

\subsection{Rotation\label{S:rot1}}
Let us first consider rotating the cube around the axis
$(0,0,0)-(1,1,1)$, counterclockwise when looking from $(1,1,1)$. This
moves the quadrilateral planes by $(n,m)\to(m,k)\to(n,k)\to(n,m)$ and
certainly does not change the consistency of the equations. Thus we
rotate cyclically the shifts $\quad\widetilde{\phantom x} \, \to \,
\widehat{\phantom x}\, \to\ \bar{\phantom x}\, \to\ \t {\phantom
  x}\quad$ accompanied with parameter change $p\,\to\,q
\,\to\,r\,\to\,p$.  Due to the cyclic convention in writing the triplet
\eqref{eq:conseqs} it means that the LHS equations become
\begin{align*}
\text{back:}\quad & Q_{12}(x,\h x,\b x,\xbh x;q,r)=0,&\\
\text{left:}\quad & Q_{23}(x,\b x,\t x,\xbt x;r,p)=0,&\\
\text{bottom:}\quad & Q_{31}(x,\t x,\h x,\xht x;p,q)=0.&
\end{align*}
This rotation does not change the roles of $c_i$ and therefore it
corresponds to the rotation of triplet codes by
\begin{equation}\label{eq:rotdef}
{\cal R}\{X,Y,Z\} = \{Z,X,Y\}.
\end{equation}
Clearly ${\cal R}^3=\mathbb I$.  Using rotation symmetry we can reduce
the number of equation triples to be analyzed, for example if we always
rotate so that $X\le Y, Z$ the remaining number of cases is $17575$
(in practice we use a slightly different method).

\subsection{Tilde-hat reflection}
In addition to rotations we need to consider some reflections.  Let us
first take the reflection across the plane defined by points $(0,0,0)
- (1,1,0) - (1,1,1), -(0,0,1)$, i.e. the tilde-hat reflection
$\quad\widetilde{\phantom x} \, \leftrightarrow \, \widehat{\phantom
  x}$, accompanied with $p \, \leftrightarrow \, q$. From the point of
view of the equations, this rule is all we need to know to get another
triplet of CAC equations. For the search process we need to know how
this and other symmetries change the triplet codes of the equations,
because we need to analyze only one triplet from the  orbit of a triplet.

Applying tilde-hat reflection on \eqref{eq:seqs} we get
\bse\label{eq:seqsr}\begin{eqnarray} \iota_1 x\h x\,c_1+\iota_2\t x \h x c_2
+\iota_3\t x\xht x\,c_3 + \iota_4x\xht x\, c_4 +\iota_5x \t x c_5
+\iota_6 \h x\xht x\,c_6 &=&0,\label{eq:sbotxr}\\ \iota_1 x\t x\,c_1
+\iota_2\t x \b x c_2 +\iota_3\b x\xbt x\,c_3 +\iota_4 x\xbt x\, c_4
+\iota_5x \b x c_5 + \iota_6\t x\xbt x\, c_6
&=&0,\label{eq:sbacxr}\\ \iota_1x\b x\,c_1 +\iota_2\b x \h x c_2
+\iota_3\h x\xbh x\,c_3 + \iota_4x\xbh x\, c_4 +\iota_5x \h x c_5 +
\iota_6 \b x \xbh x\,c_6&=&0.\label{eq:slefxr}
\end{eqnarray}\ese
Comparing this with \eqref{eq:seqs} we find that the back and left
equations get exchanged, and in addition the equations change by
exchanging the $\iota$ or $c$ subscripts $1\leftrightarrow 5$ and
$3\leftrightarrow 6$ (compare, e.g., \eqref{eq:slefxr} with
\eqref{eq:sbacx}).  Thus the tilde-hat reflection is given by the
operator $\Theta$ \bse
\begin{equation}\label{eq:thdef1}
\Theta\{X,Y,Z\} = \{\theta X,\theta Z,\theta Y\},
\end{equation}
where the $\theta$ operator changes the binary codes by
\begin{equation}\label{eq:thdef2}
\theta\{\iota_1,\iota_2,\iota_3,\iota_4,\iota_5,\iota_6\}
= \{\iota_5,\iota_2,\iota_6,\iota_4,\iota_1,\iota_3\}.
\end{equation}
\ese

The above was for tilde-hat reflection, but we can of course also have
hat-bar and bar-tilde reflections, these can be obtained using
rotations: ${\cal R} \Theta {\cal R}^2$ and ${\cal R}^2 \Theta {\cal
  R}$, respectively.

We have $\theta^2=\mathbb I$ and hence $\Theta^2=\mathbb I$,
furthermore ${\cal R}^2\Theta=\Theta{\cal R}$, $(\Theta{\cal
  R})^2=({\cal R}\Theta )^2=\mathbb I$.

\subsection{Tilde reversal}
Another important reflection is the shift reversal. Let us consider
the tilde reversal, which means reflection across the plane defined by
the points $(\tfrac12,0,0) - (\tfrac12,1,0) - (\tfrac12,1,1) -
(\tfrac12,0,1)$, or alternatively, changing all tildes to down-tildes
and then applying an overall tilde shift. This will exchange the back
and front equations, which were related by shift, and this we can
ignore.  To see how it changes the bottom and left equation we apply
the operation on \eqref{eq:seqs} and obtain
\bse\label{eq:seqst}\begin{eqnarray} \iota_1 x\t x\,c_1+\iota_2x \xht x c_2
+\iota_3\h x\xht x\,c_3 + \iota_4\t x\h x\, c_4 +\iota_5\t x \xht x c_5
+\iota_6 x\h x\,c_6 &=&0,\label{eq:sbotxt}\\ \iota_1 x\h x\,c_1
+\iota_2\h x \b x c_2 +\iota_3\b x\xbh x\,c_3 +\iota_4 x\xbh x\, c_4
+\iota_5x \b x c_5 + \iota_6\h x\xbh x\, c_6
&=&0,\label{eq:sbacxt}\\ \iota_1\t x\xbt x\,c_1 +\iota_2\xbt x x c_2
+\iota_3x\b x\,c_3 + \iota_4\t x\b x\, c_4 +\iota_5x \t x c_5 +
\iota_6 \b x \xbt x\,c_6&=&0,\label{eq:slefxt}
\end{eqnarray}\ese
Thus for the bottom equation we have $2\leftrightarrow 4$ and
$5\leftrightarrow 6$, the back equation is unchanged, and for the left
equation we have $1\leftrightarrow 3$ and $2\leftrightarrow 4$. We can write the action of the reversal operation $\cal T$ as
\bse\begin{equation}\label{eq:trevdef1} {\cal T}\{X,Y,Z\} = \{T_a
X,Y,T_b Z\}
\end{equation}
where
\begin{equation}\label{eq:trevdef2}
T_a\{\iota_1,\iota_2,\iota_3,\iota_4,\iota_5,\iota_6\}
= \{\iota_1,\iota_4,\iota_3,\iota_2,\iota_6,\iota_5\}.
\end{equation}
and
\begin{equation}\label{eq:thdef3}
T_b\{\iota_1,\iota_2,\iota_3,\iota_4,\iota_5,\iota_6\}
= \{\iota_3,\iota_4,\iota_1,\iota_2,\iota_5,\iota_6\}.
\end{equation}
\ese We have $T_aT_b=T_bT_a$, $T_a^2=\mathbb
I,\,T_b^2=\mathbb I$ and therefore ${\cal T}^2=\mathbb I$. The main
relation between $T_x$ and $\theta$ is the conjugation $\theta\,
T_a\theta=T_b$. We also have ${\cal RTR}^2=\Theta{\cal T}\Theta$, or
alternatively $(\Theta{\cal RT})^2=\mathbb I$ or ${\cal
  RTR}=\Theta{\cal T R}\Theta$, also $({\cal T}\Theta)^4=\mathbb I$.

The above was for tilde-reversal, but we can of course also have
hat-reversal and bar-reversal, these can be obtained considering:
${\cal R T R}^2$ and ${\cal R}^2 {\cal T R}$.

\subsection{Inversion}
One important transformation that preserves affine linearity of the
quad equation (up to overall factor) is the Moebius or affine linear
transformation. Now that we are dealing with homogeneous equations
only scaling $x\mapsto \text{const.}\times x$ and inversion $x\mapsto
1/x$ remain.

From \eqref{eq:seqs} we can read that inversion $x\mapsto 1/x$ implies
$1\leftrightarrow 3$, $2\leftrightarrow 4$ and $5\leftrightarrow 6$.
Similar exchanges were also seen in tilde reversal above. Indeed one
finds that $({\cal RT})^3$ implies $1\leftrightarrow 3$ and
$5\leftrightarrow 6$, but not $2\leftrightarrow 4$. Thus inversion
would be an additional symmetry operation in cases where one of
$c_2,c_4$ vanishes but not both. It will turn out that such an
asymmetric possibility arises only with solutions in which $c_2$ and
$c_4$ remain arbitrary functions, because then there is also solution
in which one of $c_2, c_4$ vanishes. Such solutions are obtained by
simple reductions and therefore we do not actually need inversion.

\subsection{Orbit of the transformation group}
By composing the basic operators of rotation $\cal R$ and two
reflections $\cal T$ and $\Theta$ in various orders one can generate
the orbit of a given initial triplet and for integrability we only
need to check one representative of each orbit.

We have already mentioned some relations between the operators (and
there may be others) and these can be used to eliminate some
combinations as redundant.  However, the full group generated by these
three operations is probably rather large, but in our case we only
need their representation when operating on three strings each
containing six zeros or ones.  It turns out that (after rotating the
triplets into some canonical form) we only have orbits of length 1, 2, 4
or 8. Below we will give the orbits separately, if they have length $>1$.

\subsection{Gauge transformation}
Finally, we can change the coefficients in the equations by a gauge
transformation of the form
\begin{align}
x_{n,m,k} \to &\phantom{\times}a(p)^{n-n_0}\, b(q)^{m-m_0}\,
c(r)^{k-k_0}\nn \\ &\times A(p)^{(n-n_0)^2}\, B(q)^{(m-m_0)^2}\,
C(r)^{(k-k_0)^2}\label{eq:fullgauge}\\ &\times {\cal
  A}(p,q)^{(n-n_0)(m-m_0)}\, {\cal B}(q,r)^{(m-m_0)(k-k_0)}\, {\cal
  C}(r,p)^{(k-k_0)(n-n_0)}\,x_{n,m,k}, \nn
\end{align}
where we have also included terms with quadratic exponent. However,
there are strong conditions on the allowed values for
$a,b,c,A,B,C,{\cal A,B,C}$ because the transformation must not break
the uniformity of the 3D lattice, i.e., the translation invariance of
the quad equation. (Note that for non-homogeneous equations the gauge
freedom would be considerably smaller, perhaps only leaving sign
changes.)

In order to check the uniformity easily we have introduced a fixed
point $n_0,m_0,k_0$, and when the gauge transformation is applied to a
given equation the result cannot depend on the fixed point, except
possibly through an overall multiplier.  The restrictions this imposes
on the parameters of the gauge transformation will depend on the particular
triplet of equations. 

We will use gauge transformations to simplify the final result, for
example by eliminating some free coefficient(s) from the equation(s),
when it simplifies the system. However, even if we can eliminate some
free function $c_i$ this way, the mere possibility that such a
function exists is by itself important for classification.

\section{The search}
Solving the equations that result from the CAC conditions cannot be fully
automatized and therefore we cannot separately solve each one of the
tentative number of 17575 cases (obtained with $X\le Y,Z$). For this
reason further computer scanning was done by checking whether the
equations obtained from the CAC condition contain any that are
monomials of the $c_j$. Since $c_j$ are all nonzero the monomial
equation cannot be satisfied and therefore that equation triplet can
be omitted. This scan turned out to effective, and it was possible to
analyze the remaining list of triplets (one representative per orbit).

For the triplet codes remaining after scanning we studied the
equations generated by CAC until we got into a contradiction with
$c_i\neq 0$, or one of the equations in the triplet factorized, or in
the positive case, all equations were solved. The search and the
results obtained are given in Appendix \ref{S:resu}. As discussed
there, we grouped the set of codes remaining after rough scanning into
blocks, and only one representative of an orbit was tested. We have
freely used gauge transformations \eqref{eq:fullgauge} and rotations
\eqref{eq:rotdef} to present the triplet equations in a nice form, for
example so that the ``side equations'' are similar.

In the next section we group and classify the result of Appendix \ref{S:resu}.

\section{Classification\label{S:class}}
The primary classification of the results is by whether or not the
triplet of equations {\em can} contain free functions of two variables
$c_i$. The gauge transformation \eqref{eq:fullgauge} can sometimes be
used to eliminate such a function but what is relevant is whether such
functions can appear (other than as overall factors).

It is also useful to observe the appearance of the $c_2,c_4$ pair,
that is, terms of the type $\t x\h x+x\xht x$. This is useful since
the pair does not get mixed up with other terms in any of the
reflections.

The results we have obtained can be grouped into four sets, each set
having a highest equation from which the remaining equations are obtained by
reductions, possibly accompanied by rotations and reflections.

\subsection{\{63,10,10\}}
For the first group the most general triplet is
\bse\label{631010}\begin{eqnarray}
x\t x\,c_1(p,q)+\t x \h x c_2(p,q)+\h x\xht x\,c_3(p,q)+
 x\xht x\, c_4(p,q)+x \h x c_5(p,q)+ \t x\xht x c_6(p,q)&=&0,\phantom{mmm}
\label{631010bot}\\
\h x \b x - x\xbh x&=&0,\label{631010bac}\\
\b x \t x - x\xbt x&=&0.\label{631010lef}
\end{eqnarray}
\ese Thus the bottom equation is a completely free homogeneous
quadratic equation while the side equations are simple.  This is
Equation \eqref{A631010} in Appendix \ref{S:A.1}.  The triply shifted
$x$ is given by
\[
\xbht x =-\frac{ c_5(p,q)x\b x\h x+c_1(p,q)x \t x \b x
  +c_2(p,q)\t x\h x\b x}{c_4(p,q)x^2+c_6(p,q)x\t x+c_3(p,q)x\h x}\,.
\]
We see that this equation has the tetrahedron property if
$c_2=c_4=0$.  The organization of the 36 sub-cases, obtained by
setting some $c_j=0$, into orbits is given in Appendix \ref{S:A.2}. The
simplest equation in this category is \eqref{101010} in Appendix
\ref{S:A.1}.

For \{63,10,10\} one cannot change the back or left equations by
gauge, but as the number of terms in the bottom equation decreases
there is more freedom in the side equations. This is illustrated by
a CAC result in the \{58,10,10\} category:
\bse\label{581010}\begin{eqnarray} \t x \h x c_2(p,q)+ x\xht
  xc_4(p,q)+x\h x\,c_5(p,q)+\t x\xht x\,c_6(p,q)
  &=&0,\label{151010bot}\\ c_2(q,r)\b x \h x + c_4(q,r) x\xbh
  x&=&0,\\ \t x \b x - x\xbt x&=&0.
\end{eqnarray}
\ese It has two free functions in the back equation. However it
is transformed into a sub-case of \eqref{631010} by the gauge
transformation
\[
x_{n,m,k}\to (-c_2(q,r)/c_4(q,r))^{(m-m_0)(k-k_0)}x_{n,m,k}\,.
\]
The possibility of extra free functions that can be gauged away
is irrelevant for MDC, but they will have information value in the
analysis of BTs in Section \ref{S:BT}.

\subsection{\{10,58,15\}}
This triplet is Equation \eqref{A105815} in Appendix \ref{S:A.2}. Two
of the equations have four free functions each
\bse\label{105815}\begin{eqnarray} \h x\t x-x\xht x=0,\\ x \b x
  c_5(q,r) + x \xbh x c_4(q,r) + \b x \h x c_2(q,r) + \xbh x \h x
  c_6(q,r)=0,\\ x \b x c_1(r,p) + x \xbt x c_4(r,p) + \b x \t x
  c_2(r,p) + \xbt x \t x c_3(r,p)=0.
\end{eqnarray}
\ese
Note that the last two equations are not connected by a cyclic variable
change but by just tilde-hat exchange (and cyclic parameter
change). The triply shifted quantity is given by
\[
\xbht x=\b x\,\frac{(c_1(r,p)x+c_2(r,p)\t x)(c_5(q,r)x+c_2(q,r)\h x)}
{(c_4(r,p)x+c_3(r,p)\t x)(c_4(q,r)x+c_6(q,r)\h x)}.
\]
The sub-cases of \eqref{105815} are listed in Appendix \ref{S:3.2.3}
and are obtainable by setting some $c_j=0$. For each of the four term
equations there are 7 acceptable sub-cases but one of them leads to the
\{63,10,10\} category, therefore there are 36 cases in this
category. Missing $c_j$ terms sometimes allow a more general bottom
equation, but that freedom can be gauged away.  TET is possible only
if $c_2(r,p)=c_4(r,p)=c_5(q,r)=c_6(q,r)=0$ or if
$c_1(r,p)=c_3(r,p)=c_2(q,r)=c_4(q,r)=0$, but both reduce
\eqref{105815} to the \{X,10,10\} category.

\subsection{\{53,15,58\}} 
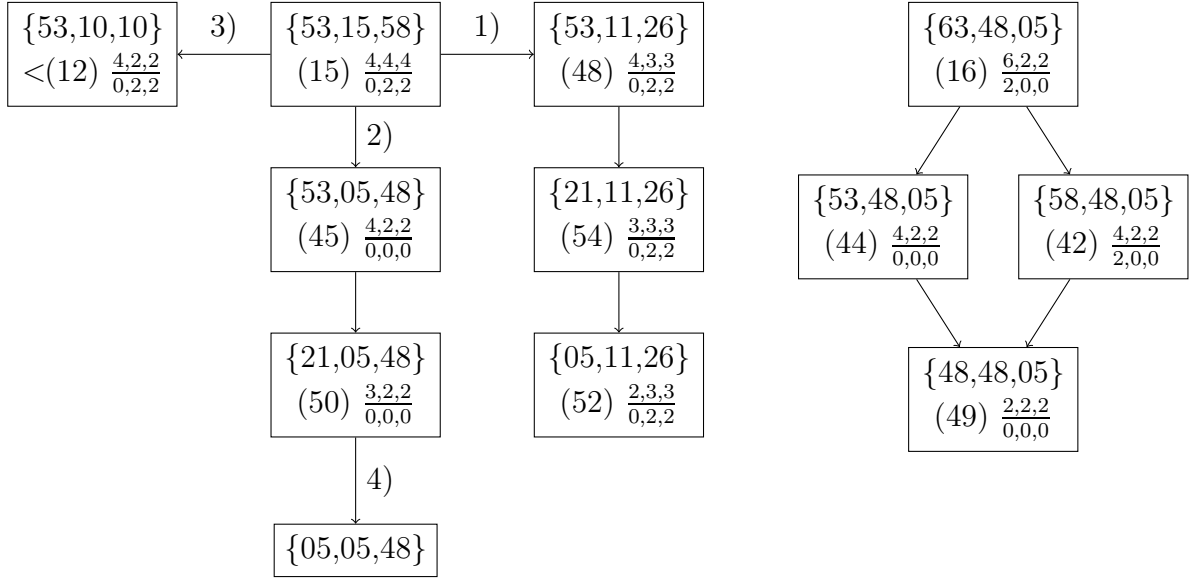
\begin{figure}
\centering
\begin{tikzpicture}[node distance=2.2cm,auto]
    \node [block] (531558) {\ynumx{531558}{4,4,4}{0,2,2}};
    \reduleftxw{531558}{531010}{3)}{4,2,2}{0,2,2}
    \redudownx{531558}{530548}{2)}{4,2,2}{0,0,0}
    \redudownx{530548}{210548}{}{3,2,2}{0,0,0}
\node [block,below of = 210548] (050548) {\{05,05,48\}};
\path[->] (210548) edge node {4)} (050548);
    \redurightx{531558}{531126}{1)}{4,3,3}{0,2,2}
    \redudownx{531126}{211126}{}{3,3,3}{0,2,2}
    \redudownx{211126}{051126}{}{2,3,3}{0,2,2}
\node [right =0cm and 1.5cm of 531126] (alku) {};
\node[block,right of= alku] (63)  {\ynumx{634805}{6,2,2}{2,0,0}};
\node [block,below left = 0.9cm and -0.8cm of 63] (53) %
{\ynumx{534805}{4,2,2}{0,0,0}};
\node [block,below right= 0.9cm and -0.8cm of 63] (58) %
{\ynumx{584805}{4,2,2}{2,0,0}};
\node [block,below left = 0.9cm and -0.8cm of 58] (48) %
{\ynumx{484805}{2,2,2}{0,0,0}};
\path[->] (63) edge node {} (53);
\path[->] (63) edge node {} (58);
\path[->] (53) edge node {} (48);
\path[->] (58) edge node {} (48);

\end{tikzpicture}
\caption{On the left some sub-case relations from \{53,15,58\}, on the
  right the reductions from \{63,48,05\}.  Each box gives the code of
  the triplet and under it on the left its equation number. The ratio
  on the right gives above the number of terms and below the number of
  $c_2,c_4$ terms.
\label{F:lim2}}
\end{figure}

Another chain of triplets that can have
free functions is obtained starting with the highest equation
{\bf{\{53,15,58\}}} or Equation \eqref{A531558} of Appendix
\ref{S:1558}.  After a gauge transformation the result can be
written as \bse\label{531558}\begin{eqnarray} c_6(p,q) \xht x \t x +
  c_3(p,q) \h x \xht x + c_1(p,q) x \t x + c_5(p,q) x \h x&=&0,\\ r(x
  \h x + \b x \xbh x) + x \xbh x + \b x \h x&=&0,\\ r(x \t x + \b x
  \xbt x) + x \xbt x + \b x \t x&=&0.
\end{eqnarray}\ese
The triply shifted variable is
\[
\xbht x=-\b x\,\frac{c_1(p,q) \t x + c_5(p,q) \h x }
{c_6(p,q) \t x + c_3(p,q) \h x}\,,
\]
and clearly has the TET property. The sub-cases obtained by setting
some $c_i=0$ change the first number of the triplet code and are
listed in Appendix \ref{S:1558}. Other types of reductions are given
in Figure \ref{F:lim2}, they are due to limits and are explained
below.

\begin{itemize}
\item[1)] \{53,11,26\} is obtained from \{53,15,58\} by scaling $\b x\to r\b x$
and then taking the leading term as $r\to 0$.
\item[2)] \{53,05,48\} is obtained from \{53,15,58\} by taking the leading
term as $r\to\infty$.
\item[3)] If $r=0$ the triplet \eqref{531558} reduces to a
sub-case \{53,10,10\} of \eqref{631010}.
\item[4)] If in \{21,05,48\} we take $c_5=0$, we get \{05,05,48\} which
  is rotated reflected \eqref{484805}.
\end{itemize}

\subsection{\{63,48,05\}}
Here the highest equation is Equation \eqref{A634805} of Appendix \ref{S:A.8}:
 \bse\label{634805}\begin{eqnarray}
 c_1(p,q)(x \t x -  \sigma_1 \sigma_3 \h x \xht x)
 + c_5(p,q) ( x \h x - \sigma_1 \sigma_2 \xht x \t x)
 + c_4(p,q) (\h x \t x - \sigma_1 x \xht x)&=&0,\\
x \b x - \sigma_3 \xbh x \h x&=&0,\\
x \b x - \sigma_2 \xbt x \t x&=&0.
\end{eqnarray}\ese
The sign $\sigma_1$ cannot be eliminated by gauge.
\[
\xbht x=\frac{x \b x (c_1(p,q) \sigma_3 \h x + c_4(p,q) x + c_5(p,q)
  \sigma_2 \t x)} {\sigma_1 \sigma_2 \sigma_3 (c_1(p,q) x \t x +
  c_4(p,q) \h x \t x + c_5(p,q) x \h x)}.
\]
It has TET if $c_4=0$, which has code \{48,48,05\}.  The diagram on
the RHS of Figure \ref{F:lim2} describes the reductions of
\eqref{634805}, they are all by setting some $c_i=0$.

\subsection{\{63,63,63\}-2, a triplet without free functions}
\begin{figure}[h]
\centering
\begin{tikzpicture}[node distance=2.2cm,auto]
    \node [block] (636363-2) {\ynumlongx{636363-2}{6,6,6}{2,2,2}};
    \redurightx{636363-2}{273159}{1)}{4,5,5}{2,2,2};
    \redudownx{273159}{261559}{2)}{3,4,5}{2,2,2};
    \redudownx{261559}{101558}{3)}{2,4,4}{2,2,2};
    \redudownlongx{636363-2}{635931-2}{4)}{6,5,5}{2,2,2};
    \reduleftlongx{636363-2}{636363-1}{6)}{6,6,6}{2,2,2};
    \redudownx{635931-2}{635815}{5)}{6,4,4}{2,2,2};
    \redudownlongx{636363-1}{635931-1}{7)}{6,5,5}{2,2,2};
\node[block,right of = {273159},node distance=3.2cm](535353)%
{\ynumx{535353}{4,4,4}{0,0,0}};
    \redudownx{535353}{534921}{9)}{4,3,3}{0,0,0};
\path[->] (636363-2)  edge [bend left=27] node  {8)} (535353);
\end{tikzpicture}
\caption{Sub-case relations, same notation as in Figure
  \ref{F:lim2}. The starting Equation \{63,63,63\}-2 \eqref{636363-2}
  is Q3$(\delta=0)$, \{63,63,63\}-1 \eqref{636363-1} is Q1$(\delta=0)$
  and \{53,53,53\} \eqref{535353} is H3$(\delta=0)$. Every reduction
  simplifies the triplet of equations and eventually we reach
  equations that can contain free functions $c_j$.
\label{F:lim1}}
\end{figure}
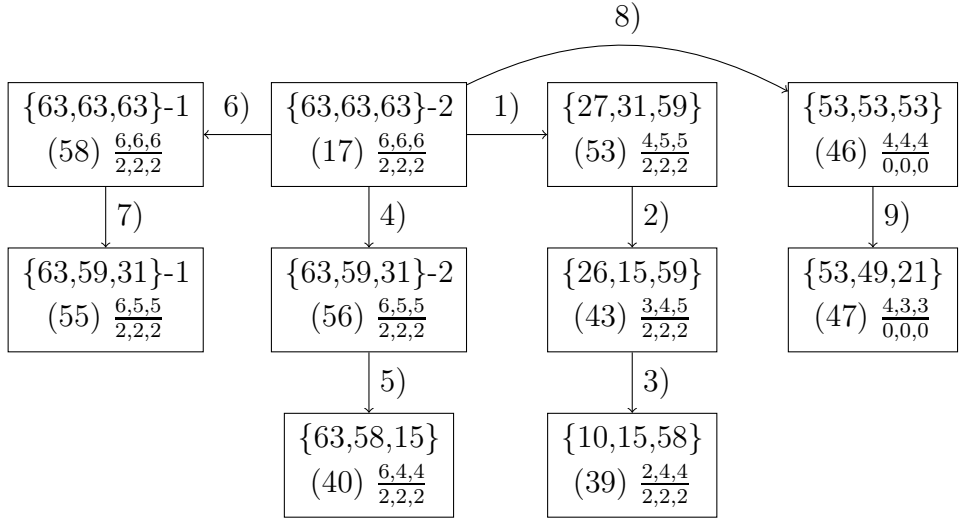

The highest level triplet of this set has code \{63,63,63\}-2, it is Equation
\eqref{A636363-2} in Appendix. It is in fact $Q3(\delta=0)$ of the ABS
list \cite{ABS03}: \bse\label{636363-2}\begin{eqnarray} (q^2 - p^2) (x
  \xht x + \h x \t x) - p (q^2-1) (x \h x + \xht x \t x) +(p^2-1) q (x
  \t x + \h x \xht x)&=&0,\\ (r^2 - q^2) (x \xbh x + \b x \h x) - q
  (r^2-1) (x \b x + \xbh x \h x) +(q^2-1) r (x \h x + \b x \xbh
  x)&=&0,\\ (p^2 - r^2) (x \xbt x + \t x \b x) - r (p^2-1) (x \t x +
  \xbt x \b x) +(r^2-1) p (x \b x + \t x \xbt x)&=&0.
\end{eqnarray}\ese
The triply shifted variable is
\[
\xbht x=\frac{\b x \h x p (q^2 - r^2) + \b x \t x q (r^2 - p^2) +
 \h x \t x r (p^2 - q^2)}{\b x  r (q^2 - p^2) + \h x q (p^2 - r^2) +
 \t x p (r^2 - q^2)}.
\]
Furthermore, all other triplets that cannot have free functions are
obtained by some limit from this equation. Figure \ref{F:lim1}
describes the chain of limits. From each orbit one element is
mentioned, its code, equation number, as well as the number of terms
in each equation and the number of $\t x\h x+x\xht x$ -type terms.
   
The limits given in Figure \ref{F:lim1} are obtained as follows:
\begin{itemize}
\item[1)] In \eqref{636363-2} scale $\t x\to -\t x/p,\,\h x\to
  -q \h x$ and change $r\to -1/r$ and then take the leading terms as
  $p\to\infty,\,q\to 0$.
\item[2)] In \eqref{273159} scale $\h x\to q\h x$ and then take the
  leading term as $q\to\infty$.
\item[3)] In \eqref{261559} scale $\t x\to p\t x$ and then take the
  leading term as $p\to\infty$. In addition replace $r\mapsto
  1/r$ and change signs by gauge.
\item[4)] In \eqref{636363-2} scale $\b x\to \b x/r$ and then take
  the leading term as $r\to\infty$. Also change signs of $p,q$.
\item[5)] In \eqref{635931-2} scale $\b x\to r\b x$ and then take
  the leading term as $r\to\infty$.
\item[6)] In \eqref{636363-2} let $p\mapsto 1+\epsilon p,\,q\mapsto
  1+\epsilon q,\,r\mapsto 1+\epsilon r$ and take the leading term as
  $\epsilon\to0$.
\item[7)] In \eqref{636363-1} scale $\b x\to \b x/r,\ \t x\to -\t x,\,
  \h x\to-\h x$ and then take leading term in $r\to\infty$.
\item[8)] In \eqref{636363-2} take the leading term (linear) as $p,q,r\to 0$.
\item[9)] In  \eqref{535353}  scale $\b x\to r\b x$
and then take the leading terms as $r\to 0$.
\end{itemize}
In this way all equations to which free functions of two variables
cannot be introduced by a gauge are reductions of $Q3(\delta=0)$
\eqref{636363-2}. 

It should be noted, however, that in Figure \ref{F:lim1} the bottom
box \{10,15,58\} \eqref{101558} is an equation that does allow free
functions, and this happens also on further reductions from the
other bottom boxes: From \{63,58,15\} \eqref{635815} the limit $p,q\to
0$ leads to \{53,10,10\} while $p,q\to\infty$ leads to \{53,48,05\}
\eqref{534805}; from \{10,15,58\} \eqref{101558} the limit $r\to$
leads to \{10,05,10\} while $r\to\infty$ leads to \{10,10,48\}, which
are in the same orbit; from \{53,49,21\} \eqref{534921} the limit
$p,q\to\infty$ leads to \{53,48,05\} \eqref{534805}. All of these
further limits yield equations which allow free functions. It should
be noted also that some potential limits actually lead to factorizable
equations, for example if in \{63,59,31\}-1 \eqref{635931-1} one takes
$p,q\to 0$ or if in \{63,58,15\} \eqref{635815} one takes $p\mapsto
1+\epsilon p,\,q\mapsto 1+\epsilon q,\,r\mapsto 1+\epsilon r$ and then
take the leading term as $\epsilon\to0$.

We also note that \eqref{636363-2} still allows gauge transformations
of the form
\begin{align}
x_{n,m,k} \to &\phantom{\times}a(p)^{n-n_0}\, b(q)^{m-m_0}\,
c(r)^{k-k_0}\, \sigma_1^{(n-n_0)^2}\, \sigma_2^{(m-m_0)^2}\,
\sigma_3^{(k-k_0)^2}
\end{align}
where $\sigma_j^2=1$. This changes the bottom equation into
\[
 [x \xht x + \h x \t x] (q^2 - p^2)-[x \h x a(p)^{-1} + \xht x \t x
   a(p)] \sigma_1 p (q^ 2 - 1) + [x \t x b(q)^{-1} + \h x \xht x b(q)]
 \sigma_2 q (p^2 - 1)=0,
\]
and corresponding results hold for the other equations.

\subsection{Comparison with Boll's results}
As was mentioned before, Boll has done a classification of triplets of
CAC equations in \cite{Boll2011,Boll2012JNMP,Boll2012}. His setup was
both more general (allowing flips between opposing sides and not
restricting to homogeneous quadratic equations) and less general (by
requiring the tetrahedron property).  In \cite{Boll2011} the results
were collected into Theorems 3.4-3.11. Restricting those equations to
quadratic irreducible equations without flips implies some
restrictions on the parameters, and yield the following:
\begin{enumerate}
  \item For (3.6),(3.4) we must take $\delta=0,\epsilon=0$, then it gives
    \eqref{535353}, i.e. H3$(\delta=0)$. If $\gamma=0$ we get \eqref{534805}
  \item For (3.10),(3.7) we must take $\delta=0$, then it gives
    \{47,63,62\}, which is reflected rotated \eqref{635931-2}. If
    furthermore $\epsilon=0$ we get rotated \eqref{635815}.
  \item For (3.14),(3.11) we must take $\delta=0,\,\epsilon=0$, then
    it gives \eqref{531558}. If furthermore we take $\gamma=0$ we get
    \eqref{534805}. (Also for $\gamma^2=1$ there are no flips but then
    some equations factorize.)
  \item In (3.19-20) one additional parameter dependence is missing in
    $B$, this is corrected in \cite{Boll2012JNMP}. In the corrected
    form it is necessary to take $\delta_3=0$, which yields reflected
    \eqref{273159}. If after this we take $\delta_1=0$ or $\delta_2=0$
    we get a reflected forms of \eqref{261559} and if both
    coefficients vanish we get \eqref{101558}.
  \item For (3.21-22) (of \cite{Boll2011}, not \cite{Boll2012JNMP}) we
    must take $\delta_1=0$ or change $\delta x_1x_2$ into $\delta
    x_1x$, which is again a special case of \eqref{531558}. If $\alpha^2=1$
    some equations factorize.
  \item For (3.29-30) we must take $\delta_2=0$, then it gives
    reflected rotated \eqref{534921}. If furthermore $\delta_1=0$ we
    get rotated \eqref{534805}.
  \item For (3.31-32) we must take $\delta_1=\delta_2=0$, then it
    gives rotated \eqref{530548}. In that case all coefficients in $B$
    of (3.31) can in fact be free. If in (3.31) we take $\alpha=0$ we
    get rotated \eqref{484805}.
  \item For (3.33-34) we must take $\delta_1=0$, then it gives
    reflected \eqref{531126}. In $B$ all coefficients can be free.
  \item For (3.41-42) we must take $\delta_1=0$, then it corresponds
    to \{42,21,14\}, a rotated form of which appears in the orbit of
    \eqref{211126}. Next if $\delta_3=0$ we get rotated reflected
    \eqref{051126} and if $\delta_2=0$ we get subcase \{53,10,10\} of
    \eqref{631010}.
  \item For (3.43-44) we must take $\delta_1=\delta_2=0$, then it
    becomes rotated form of \eqref{210548}. (Note that in the formula
    for $C$ the first variable must be $x_{13}$, not $x_{12}$.)
\end{enumerate}
Thus all homogeneous quadratic shift invariant equations in Boll's
classification \cite{Boll2011} are also included in our
classification. However, there is one result we could not identify
among Boll's results, namely \eqref{635931-1}, which is related to Q1
the same way that \eqref{635931-2} is related to Q3, and
\eqref{535353} is related to H3.

\subsection{Triplets with two term equations}
As another selection of our results we would like to collect here the
simplest CAC triplets, those for which all equations have just two
terms. Below on the LHS they are given in the general form after all
equations from the CAC condition have been solved, and then on the RHS
we present the simplest form obtainable by a gauge transform.

\begin{itemize}
\item {\bf\{10,10,10\}}
\begin{equation}\left\{\begin{array}{rcl}
c_2(p,q) \t x\h x + c_4(p,q) x \xht x&=&0,\\
c_2(q,r) \h x\b x + c_4(q,r) x \xbh x&=&0,\\
c_2(r,p) \b x\t x + c_4(r,p) x \xbt x&=&0,
\end{array}\right.\quad\rightarrow\quad\left\{
\begin{array}{rcl}
\t x\h x - x \xht x&=&0,\\
\h x\b x - x \xbh x&=&0,\\
\b x\t x - x \xbt x&=&0.
\end{array}\right.
\end{equation}

\item {\bf\{5,10,10\}} or \{48,10,10\} by tilde-hat reflection
\begin{equation}\left\{\begin{array}{rcl}
c_1(p,q) x\t x + c_3(p,q) \h x \xht x&=&0,\\
\h x\b x + \sigma x \xbh x&=&0,\\
c_2(r,p) \b x\t x + c_4(r,p) x \xbt x&=&0,
\end{array}\right.\quad\rightarrow\quad\left\{
\begin{array}{rcl}
c_1(p,q)x\t x + c_3(p,q) \h x \xht x&=&0,\\
\h x\b x - x \xbh x&=&0,\\
\b x\t x - x \xbt x&=&0.
\end{array}\right.
\end{equation}

\item {\bf\{10,48,5\}}
\begin{equation}\left\{\begin{array}{rcl}
 \t x\h x + \sigma x \xht x&=&0,\\
c_5(q,r) x\b x + c_6(q,r) \h x \xbh x&=&0,\\
c_1(r,p) x\b x + c_3(r,p) \t x \xbt x&=&0,
\end{array}\right.\quad\rightarrow\quad\left\{
\begin{array}{rcl}
\t x\h x - x \xht x&=&0,\\
c_5(q,r) x\b x + c_6(q,r) \h x \xbh x&=&0,\\
c_1(r,p) x\b x + c_3(r,p) \t x \xbt x&=&0.
\end{array}\right.
\end{equation}

\item {\bf\{5,5,48\}} and \{48,5,48\} by tilde-hat reflection
\begin{equation}\left\{\begin{array}{rcl}
c_1(p,q) x\t x + c_3(p,q) \h x \xht x&=&0,\\
c(r) x\h x + \b x \xbh x&=&0,\\
\sigma c(r)  x\t x +  \b x \xbt x&=&0,
\end{array}\right.\quad\rightarrow\quad\left\{
\begin{array}{rcl}
c_1(p,q) x\t x + c_3(p,q) \h x \xht x&=&0,\\
 x\h x - \b x \xbh x&=&0,\\
 x\t x - \b x \xbt x&=&0.
\end{array}\right.
\end{equation}

\end{itemize}

These cases are easily distinguished by the number of $c_2,c_4$ terms,
which have also been rotated into a particular position. These
triplets can be obtained from several results listed in the Appendix
by various limits.

\subsection*{Summary:} The triplets that were found using CAC can be 
divided into two groups: 1) Those that cannot have free functions of two
lattice parameters are all reductions of $Q3(\delta=0)$ of the ABS
list \cite{ABS03}. 2) The triplets that {\em can} have free functions of
two lattice parameters can be organized into 4 sets, with the highest
triplets of the set being \{63,10,10\}, \{53,15,58\}, \{63,48,05\}, or
\{10,58,15\}.

\section{B\"acklund transformations and Lax pairs}
The equations that were found using CAC will next be studied from the
point of view of their B\"acklund transformations and Lax pairs.

\subsection{General setup}
Let us fix the equations and some concepts.  We have six
equations, which we can arrange as follows:
\bse\label{eq:cacbteqs}\begin{align} Q_{12}(y,{\t y},{\h y},&{\h{\t
      y}};p,q)=0,\label{eq:BTtop}\\ Q_{23}(x,\h x,y,{\h
    y};q,r)=0,\quad &Q_{23}(\t x,\h{\t x},{\t y},{\h{\t
      y}};q,r)=0\label{eq:BTside1}\\ Q_{31}(x,y,\t x,{\t
    y};r,p)=0,\quad & Q_{31}(\h x,{\h y},\h{\t x},{\h{\t
      y}};r,p)=0,\label{eq:BTside2}\\ Q_{12}(x,\t x,\h x,&\h{\t
    x};p,q)=0.\label{eq:BTbottom}
\end{align}\ese
This is the same set as \eqref{eq:conseqs}, except that we have named $\b
x=y$, which has no effect on the algebra.  We see that the top
equation \eqref{eq:BTtop} depends only on the $y$ variables, the
bottom equation \eqref{eq:BTbottom} only on the $x$ variables, while
the side equations \eqref{eq:BTside1},\eqref{eq:BTside2} depend on two
$x$ and on two $y$ variables.

\begin{definition}[B\"acklund transform of equations (eqBT)]\label{D:2}
Consider equations \eqref{eq:BTside1} and \eqref{eq:BTside2} and solve
for $\h y$ and $\t y$ from their LHS. Then from the RHS of these
equations we can solve for $\xht y$ in two ways. If these two results
are automatically equal, we say that the {\em eqBT is trivial}, i.e.,
it fails. If the results are different we factorize the numerator of
their difference and take all factors that only depend on $x,\t x,\h
x,\h{\t x}$. If there is a unique factor that is also an acceptable
equation the we say that the {\em eqBT is strong} and the
generated equation is called $Q_{12}$ \eqref{eq:BTbottom}. If there are several
acceptable factors we say that the {\em eqBT is weak}. The same
process can be done by eliminating the $x$ functions, leading
(possibly) to an equation in $y$.
\end{definition}

Note that if the system \eqref{eq:cacbteqs} has the CAC property then
the above mentioned two results for $\xht y$ will be the same either
automatically or modulo $Q_{12}$.

\begin{definition}[B\"acklund transform of solutions (solBT)]
Suppose $x$ is a solution of \eqref{eq:BTbottom}. If we can solve for
$y$ from \eqref{eq:BTside1} and \eqref{eq:BTside2} and that solution
solves \eqref{eq:BTtop} then we have a BT transforming solutions. If
$x=y$ then the transformation is trivial.
\end{definition}

Since we are only interested in the equations and their relationships
we only use eqBT as described in Definition \ref{D:2}.

Before doing any computations we can already make some statements
about the triplets that we have found. The starting setup was that all
coefficients in the triplet of equations were free functions depending
on the two lattice parameters associated with that quadrilateral, and
then from the CAC we got conditions due to which some two-variable
functions were fixed. However, we have seen that sometimes even if a
triplet is CAC it may still contain arbitrary functions $c_i$ (other
than overall factors).

If we now look at this from the point of view of eqBT, it is clear
that an equation with a free function cannot be generated by an eqBT
composed of the remaining pair of equations, because of different
dependence on lattice parameters. For example, when the side equations
only contain $c_j(q,r)$ and $c_j(r,p)$ they cannot be used to
construct arbitrary free $c_j(p,q)$. Furthermore, since gauge
transformations have no effect on the eqBT, it turns out that the
important property is the {\em possibility} of free functions (other than
overall multipliers). For example for \eqref{101010} BT produces
nothing and this can be predicted from the fact that by gauge freedom
each equation can have free functions as shown in
\eqref{101010-free}. In the case of the triplet \eqref{581010} the
bottom and back equations may contain free functions and therefore a
eqBT can at most produce the back equation.

In comparing eqBT and CAC we note that for CAC there is no fundamental
difference in the roles of the equations, while for eqBT the middle
equations \eqref{eq:BTside1}, \eqref{eq:BTside2} form the
transformation using which we should be able to generate top and
bottom equations.  In a closer analysis we observe that for CAC and
eqBT the computations start the same way: If we take $x,\,\t
x,\,\h x,\,y$ as initial values we can solve for $\h y$ in terms of
$x,\,\h x,\, y$ from the back equation and $\t y$ in terms of $x,\,\t
x,\, y$ from the left equation. Then solving for $\xht y$ from the
right equation we have three equations remaining, bottom equation,
which depends on $x,\,\t x,\,\h x,\,\xht x$, and top and front
equations which now depend on $x,\,\t x,\,\h x,\,\xht x,\,y$ At this
point the CAC and BT methods diverge.
\begin{itemize}
\item In the next step for CAC we would solve
for $\xht x$ from the bottom equation and when it is substituted into
the two remaining equations they should vanish identically. 
\item In eqBT we are only working with $Q_{23}$ and $Q_{31}$, and
  therefore only have the front equation remaining. It should now
  factorize with one factor being the bottom equation, the other
  factor possibly containing $y$.
\end{itemize}
In practice the methods differ when the front equation vanishes {\em
  before} substituting $\xht x$, in which case we clearly cannot
generate the bottom equation.

We note also that in \eqref{eq:cacbteqs} the roles of $x$ and $y$ are
symmetric and therefore the same arguments can be used to generate the
top equation. Furthermore, if we instead rename $\h x=y$ in
\eqref{eq:conseqs} we get a set from which the above procedure can be
used, mutatis mutandis, for generating left or right equations, and if
we take $\t x=y$ we can generate back and front equations.

\subsection{Computations for eqBT}
In order to further analyze the eqBT, it is useful to develop the formulae
to some extent without specifying the form of the $Q$-polynomials.
For this purpose it is often useful to isolate one of the variables
of $Q$, but since $Q$ is multilinear this is easy. We can write, for example,
\[
Q_{12}(x,\t x,\h x,\h{\t x};p,q)=Q_{12}(x,\t x,\h x,\bullet;p,q)\h{\t x}
+Q_{12}(x,\t x,\h x,0;p,q)
\]
where
\[
Q_{12}(x,\t x,\h x,\bullet;p,q):=\frac{\partial}{\partial \h{\t
    x}}Q_{12}(x,\t x,\h x,\h{\t x};p,q) = Q_{12}(x,\t x,\h
x,1;p,q)-Q_{12}(x,\t x,\h x,0;p,q)
\]
Using this notation we can solve the LHS equations of
\eqref{eq:BTside1}, \eqref{eq:BTside2}, \bse\label{eq:cac-solved}
\begin{eqnarray}
{\h y}=-\frac{Q_{23}(x,\h x,y,0;q,r)}{Q_{23}(x,\h x,y,\bullet;q,r)},\\
{\t y}=-\frac{Q_{31}(x,y,\t x,0;r,p)}{Q_{31}(x,y,\t x,\bullet;r,p)}.
\end{eqnarray}
\ese
and then from the RHS equations we can solve
\bse\begin{eqnarray}
{\h{\t{ y}}}_{23}&=&-\frac{Q_{23}(\t x,\h {\t x},{\t y},0;q,r)}
{Q_{23}(\t x,\h{\t x},{\t y},\bullet;q,r)}\nn\\
&=&\frac
{Q_{23}(\t x,\h {\t x},\bullet,0;q,r)Q_{31}( x,{ y},{\t x},0;r,p)
-Q_{23}(\t x,\h {\t x},0,0;q,r)Q_{31}( x,{y},{\t x},\bullet;r,p)}
{Q_{23}(\t x,\h {\t x},\bullet,\bullet;q,r)Q_{31}( x,{ y},{\t x},0;r,p)
-Q_{23}(\t x,\h {\t x},0,\bullet;q,r)Q_{31}( x,{y},{\t x},\bullet;r,p)},
\phantom{mmm}
\label{eq:cac-fin23}\\
{\h {\t y}}_{31}&=&-\frac{Q_{31}(\h x,{\h y},\h{\t x},0;r,p)}
{Q_{31}(\h x,{\h  y},\h{\t x},\bullet;r,p)}\nn\\
&=&\frac
{Q_{31}(\h x,\bullet,\h{\t x},0;r,p)Q_{23}(x,\h x,y,0;q,r)-
Q_{31}(\h x,0,\h{\t x},0;r,p)Q_{23}(x,\h x,y,\bullet;q,r)}
{Q_{31}(\h x,\bullet,\h{\t x},\bullet;r,p)Q_{23}(x,\h x,y,0;q,r)-
Q_{31}(\h x,0,\h{\t x},\bullet;r,p)Q_{23}(x,\h x,y,\bullet;q,r)}
.\label{eq:cac-fin31}
\end{eqnarray}\ese
Let us furthermore expand these in $y$ since it cannot appear in final
equation. We have
\bse\label{eq:v23v31}\begin{eqnarray}
{\h{\t{ y}}}_{23}&=&-\frac{W_1(x,\t x,\xht x)\,y+W_2(x,\t x,\xht x)}
{W_3(x,\t x,\xht x)\,y+W_4(x,\t x,\xht x)},\\
{\h{\t{ y}}}_{31}&=&-\frac{Z_1(x,\h x,\xht x)\,y+Z_2(x,\h x,\xht x)}
{Z_3(x,\h x,\xht x)\,y+Z_4(x,\h x,\xht x)},
\end{eqnarray}\ese
where
\begin{eqnarray*}
W_1(x,\t x,\xht x)&=&
Q_{23}(\t x,\h {\t x},\bullet,0;q,r)Q_{31}( x,\bullet,{\t x},0;r,p)-
Q_{23}(\t x,\h {\t x},0,0;q,r)Q_{31}( x,\bullet,{\t x},\bullet;r,p),\\
W_2(x,\t x,\xht x)&=&
Q_{23}(\t x,\h {\t x},\bullet,0;q,r)Q_{31}( x,0,{\t x},0;r,p)-
Q_{23}(\t x,\h {\t x},0,0;q,r)Q_{31}( x,0,{\t x},\bullet;r,p),\\
W_3(x,\t x,\xht x)&=&
Q_{23}(\t x,\h {\t x},\bullet,\bullet;q,r)Q_{31}( x,\bullet,{\t x},0;r,p)-
Q_{23}(\t x,\h {\t x},0,\bullet;q,r)Q_{31}( x,\bullet,{\t x},\bullet;r,p),\\
W_4(x,\t x,\xht x)&=&
Q_{23}(\t x,\h {\t x},\bullet,\bullet;q,r)Q_{31}( x,0,{\t x},0;r,p)-
Q_{23}(\t x,\h {\t x},0,\bullet;q,r)Q_{31}( x,0,{\t x},\bullet;r,p),\\
Z_1(x,\h x,\xht x)&=&
Q_{31}(\h x,\bullet,\h{\t x},0;r,p)Q_{23}(x,\h x,\bullet,0;q,r)-
Q_{31}(\h x,0,\h{\t x},0;r,p)Q_{23}(x,\h x,\bullet,\bullet;q,r),\\
Z_2(x,\h x,\xht x)&=&
Q_{31}(\h x,\bullet,\h{\t x},0;r,p)Q_{23}(x,\h x,0,0;q,r)-
Q_{31}(\h x,0,\h{\t x},0;r,p)Q_{23}(x,\h x,0,\bullet;q,r),\\
Z_3(x,\h x,\xht x)&=&
Q_{31}(\h x,\bullet,\h{\t x},\bullet;r,p)Q_{23}(x,\h x,\bullet,0;q,r)-
Q_{31}(\h x,0,\h{\t x},\bullet;r,p)Q_{23}(x,\h x,\bullet,\bullet;q,r),\\
Z_4(x,\h x,\xht x)&=&
Q_{31}(\h x,\bullet,\h{\t x},\bullet;r,p)Q_{23}(x,\h x,0,0;q,r)-
Q_{31}(\h x,0,\h{\t x},\bullet;r,p)Q_{23}(x,\h x,0,\bullet;q,r).
\end{eqnarray*}
We use the symbol $\oeq$ for equalities that should hold modulo an
acceptable equation. For a set of equations the equality should be
modulo the {\em same} equation, which we call $Q_{12}$.

Since $Q_{12}$ does not contain $y$, the B\"acklund condition ${\h{\t{
      y}}}_{23}\,\oeq\,{\h{\t{ y}}}_{31}$ implies three equations
\bse\label{eq:bt-gen}\begin{align} y^2:& & W_1 Z_3- W_3
  Z_1&\oeq0\label{eq:bt-gen2},\\ y^1:& &
  W_1Z_4+W_2Z_3-W_4Z_1-W_3Z_2&\oeq0,\label{eq:bt-gen1}\\ y^0:& &
  W_2Z_4-W_4Z_2&\oeq0.\label{eq:bt-gen0}
\end{align}
\ese For a genuine eqBT these equations should not all be satisfied
automatically but rather for at least one equation the LHS should
factor with an acceptable equation as a factor.

\subsection{Deriving the Lax condition}
It is interesting to compare eqBT with conditions derived from a Lax
pair.  The starting point is still \eqref{eq:cac-solved}, but now we
replace $y\to f/g$. Thus for example
\begin{eqnarray*}
\frac{\h f}{\h g}&=&-\frac{Q_{23}(x,\h x,f/g,0;q,r)}
{Q_{23}(x,\h x,f/g,\bullet;q,r)}\\
&=&-\frac{Q_{23}(x,\h x,\bullet,0;q,r)\,f+Q_{23}(x,\h x,0,0;q,r)\,g}
{Q_{23}(x,\h x,\bullet,\bullet;q,r)\,f+Q_{23}(x,\h x,0,\bullet;q,r)\,g},
\end{eqnarray*}
that is
\bse\begin{equation}
\begin{pmatrix}\h f\,\\ \h g\end{pmatrix}=
{\cal M}\begin{pmatrix}f\\ g\end{pmatrix},\quad
{\cal M}=\mu(x,\h x)\begin{pmatrix}
-Q_{23}(x,\h x,\bullet,0;q,r) &-Q_{23}(x,\h x,0,0;q,r)\\
Q_{23}(x,\h x,\bullet,\bullet;q,r) & Q_{23}(x,\h x,0,\bullet;q,r)
\end{pmatrix},
\end{equation}
where $\mu$ is a separation factor. Similarly
\begin{equation}
\begin{pmatrix}\t f\,\\ \t g\end{pmatrix}=
{\cal L}\begin{pmatrix}f\\ g\end{pmatrix},\quad
{\cal L}=\lambda(x,\t x)\begin{pmatrix}
-Q_{31}(x,\bullet,\t x,0;r,p) &-Q_{31}(x,0,\t x,0;r,p)\\
Q_{31}(x,\bullet,\t x,\bullet;r,p) & Q_{31}(x,0,\t x,\bullet;r,p)
\end{pmatrix}.
\end{equation}\ese
Then the commutativity condition
\[
\begin{pmatrix}\t f\,\\ \t g\end{pmatrix}\!\widehat{\phantom{\frac11}}
\oeq\begin{pmatrix}\h f\,\\ \h g\end{pmatrix}\!\widetilde{\phantom{\frac11}}
\]
implies
\[
\widehat{\cal L}\,{\cal M}\oeq\widetilde{\cal M}\,{\cal L}.
\]
The entries of this matrix equation are
\bse\label{eq:lax-eqs}
\begin{align}
  (1,1):& & \lambda(x,\t x)\,\mu(\t x,\xht x) \, W_1&\oeq
\lambda(\h x,\xht x)\,\mu(x,\h x) \, Z_1,\\
  (1,2):& & \lambda(x,\t x)\,\mu(\t x,\xht x) \, W_2&\oeq
\lambda(\h x,\xht x)\,\mu(x,\h x) \, Z_2,\\
  (2,1):& & \lambda(x,\t x)\,\mu(\t x,\xht x) \, W_3&\oeq
\lambda(\h x,\xht x)\,\mu(x,\h x) \, Z_3,\\
  (2,2):& & \lambda(x,\t x)\,\mu(\t x,\xht x) \, W_4&\oeq
\lambda(\h x,\xht x)\,\mu(x,\h x) \, Z_4,
\end{align}
\ese
using the previous notation. By taking ratios we get
\[
W_j/W_k\oeq Z_j/Z_k,\quad\forall j,k,
\]
that is, polynomial equations
\begin{equation}\label{eq:WZlax}
W_j\, Z_k- W_k\, Z_j\oeq 0,\quad\forall j,k.
\end{equation}
These equations should either vanish, or factorize with the desired
equation as a factor.

When comparing \eqref{eq:WZlax} with \eqref{eq:bt-gen} it is easy to
verify that \eqref{eq:WZlax} implies \eqref{eq:bt-gen}. However, in the
other direction one finds that in addition to
\[
W_1/W_3\oeq Z_1/Z_3,\quad W_2/W_4\oeq Z_2/Z_4,
\]
from \eqref{eq:bt-gen2}, \eqref{eq:bt-gen0},  one finds
\[
(W_4/W_3-Z_4/Z_3)(W_1W_4-W_3W_4)\oeq 0,
\] 
from \eqref{eq:bt-gen1} after eliminating $Z_1,Z_2$. This suggests a
possible new and different type of solution with $W_1W_4-W_3W_4=0$.
This, however, does not work: for the equations under consideration
the expression $W_1W_4-W_3W_4$ never vanishes, and furthermore it
cannot be used to generate a quad equation, as it depends only on 3 of
the 4 corner variables. Thus we conclude that eqBT and Lax generate the
same equations or fail together.

\subsection{BT for the results obtained\label{S:BT}}
Above we have described how the eqBT computations lead to equations
\eqref{eq:bt-gen} which should not hold automatically, but only
after using $Q_{12}=0$, in other words, ${\h{\t{ y}}}_{23}-{\h{\t{
      y}}}_{31}$ should factorize, with $Q_{12}(x,\t x ,\h x,\xht
x;p,q)$ as one of the factors. We can of course do these computations
for all three choices of side equations: 
\begin{enumerate}
\item back-left-front-right equations producing bottom and top  equations,
\item bottom-back-top-front equations producing left and right  equations,
\item bottom-left-top-right equations producing back and front equation.
\end{enumerate}
We have done these computations for all equations found to have CAC
(one representative per orbit) as listed in Appendix \ref{S:resu}. The
results are given in Tables 1-3.  In the table a ``0'' means the
corresponding eqBT produces nothing, i.e., the equations
\eqref{eq:bt-gen} are satisfied automatically, a ``1'' means the
desired equation is uniquely produced. The remaining cases have a
``2'' indicating that the polynomial produced by the BT has two
factors that can be taken as genuine equations (i.e. they are
acceptable by Definition \ref{D:1} and do not depend on the auxiliary
variable $y$). A plain ``2'' means the two possibilities differ only
by a sign, ``$2c^n$'' means that the extra equation has degree $n$
dependence on the $c_j$, while 2* stands for the cases where the two
equations are essentially different and the extra equation is a
version of $\t x\h x-x\xht x$. Finally, if one of the resulting
equations is not quadratic its degree is given as subscript.  


\begin{table}
\begin{center}
\begin{subtable}[b]{7.7cm}
\begin{tabular}{|c|c|c|c|c|c|c|c|}
\hline
Eqn &{\rotatebox[origin=c]{90}{{\scriptsize Top}}} & 
{\rotatebox[origin=c]{90}{{\scriptsize Bottom}}} &
{\rotatebox[origin=c]{90}{{\scriptsize Left}}} &
{\rotatebox[origin=c]{90}{{\scriptsize Right}}} &
{\rotatebox[origin=c]{90}{{\scriptsize  Back}}} &
{\rotatebox[origin=c]{90}{{\scriptsize  Front}}} & TET\\
\hline
63,10,10   & 0 & 0 & 1 & 1 & 1 & 1 & no \\
62,10,10   & 0 & 0 & 1 & 1 & 1 & 1 & no \\
58,10,10   & 0 & 0 & 2$c^2$ & 2$c^2$ & 0 & 0 & no \\
42,10,10   & 0 & 0 & 1 & 1 & 0 & 0 & no \\
10,10,10   & 0 & 0 & 0 & 0 & 0 & 0 & no \\ \hline
43,10,10   & 0 & 0 & 1 & 1 & 1 & 1 & no \\ \hline
61,10,10   & 0 & 0 & 1 & 1 & 1 & 1 & no \\
60,10,10   & 0 & 0 & 1 & 1 & 1 & 1 & no \\
56,10,10   & 0 & 0 & 2$c^1$ & 2$c^1$ & 0 & 0 & no \\ \hline
53,10,10   & 0 & 0 & 1 & 1 & 1 & 1 & yes \\
52,10,10   & 0 & 0 & 1 & 1 & 1 & 1 & yes \\
05,10,10   & 0 & 0 & 0 & 0 & 2 & 2 & yes \\
\hline 
\end{tabular}\subcaption{eqBT for equations of type \{X,10,10\}. 
}\end{subtable}%
\hspace{0.6cm}\begin{subtable}[b]{7.7cm}\begin{tabular}{|c|c|c|c|c|c|c|c|}
\hline
Eqn &{\rotatebox[origin=c]{90}{{\scriptsize Top}}} & 
{\rotatebox[origin=c]{90}{{\scriptsize Bottom}}} &
{\rotatebox[origin=c]{90}{{\scriptsize Left}}} &
{\rotatebox[origin=c]{90}{{\scriptsize Right}}} &
{\rotatebox[origin=c]{90}{{\scriptsize  Back}}} &
{\rotatebox[origin=c]{90}{{\scriptsize  Front}}} & TET\\
\hline
10,58,15   & 2$c^4$ & 2$c^4$ & 0 & 0 & 0 & 0 & no \\
10,58,07   & 2$c^3$ & 2$c^3$ & 0 & 0 & 0 & 0 & no \\
10,58,11   & 2$c^3$ & 2$c^3$ & 0 & 0 & 0 & 0 & no \\
10,26,11   & 1 & 1 & 0 & 0 & 0 & 0 & no \\
10,50,11   & 2$c^2$ & 2$c^2$ & 0 & 0 & 0 & 0 & no \\
10,50,07   & 1 & 1 & 0 & 0 & 0 & 0 & no \\
10,58,05   & 2$c^2$ & 2$c^2$ & 0 & 0 & 0 & 0 & no \\
10,56,05   & 1 & 1 & 0 & 0 & 0 & 0 & no \\
10,26,05   & 2$c^1$ & 2$c^1$ & 0 & 0 & 0 & 0 & no \\
10,48,05   & 0 & 0 & 0 & 0 & 0 & 0 & no \\
\hline 
\end{tabular}\subcaption{eqBT for sub-cases of \{10,58,15\}.}
\end{subtable}
\end{center}
\caption{Results of eqBT for two classes of equations that can have
  free functions. ``2'' in a column means BT produces two
  possibilities, `` 2$c^n$ '' that the non-standard alternative has
  polynomial coefficients of degree $n$ in $c_i$. Only one
  representative of each orbit is listed. In the last column we
  indicate whether the triplet has the tetrahedron property.}
\end{table}

\begin{table}
\begin{center}
\begin{tabular}{|c|c|c|c|c|c|c|c|}
\hline
Eqn &{\rotatebox[origin=c]{90}{{\scriptsize Top}}} & 
{\rotatebox[origin=c]{90}{{\scriptsize Bottom}}} &
{\rotatebox[origin=c]{90}{{\scriptsize Left}}} &
{\rotatebox[origin=c]{90}{{\scriptsize Right}}} &
{\rotatebox[origin=c]{90}{{\scriptsize  Back}}} &
{\rotatebox[origin=c]{90}{{\scriptsize  Front}}} & TET\\
\hline
53,15,58 \eqref{531558} & 0 & 0 & 1 & 1 & 1 & 1 & yes \\
53,05,48 \eqref{530548} & 0 & 0 & 1 & 1 & 1 & 1 & yes \\
21,05,48 \eqref{210548} & 0 & 0 & 1 & 1 & 1 & 1 & yes \\
\hline
53,11,26 \eqref{531126}  & 0 & 0 & 1 & 1 & 1 & 1 & yes \\
21,11,26 \eqref{211126}  & 0 & 0 & 1 & 1 & 1 & 1 & yes \\
05,11,26 \eqref{051126}  & 0 & 0 & 1 & 1 & $2^*$ & $2^*$ & yes \\
\hline
63,48,05 \eqref{634805}  & 0 & 0 & 1 & 1 & 1 & 1 & no \\
58,48,05 \eqref{584805} & 0 & 0 & 2$c^1$ & 2$c^1$ & 0 & 0 & no \\
53,48,05 \eqref{534805} & 0 & 0 & 1 & 1 & 1 & 1 & yes \\
48,48,05 \eqref{484805} & 0 & 0 & 2 & 2 & 0 & 0 & yes\\
\hline
10,15,58 \eqref{101558} & 0 & 0 & 2* & 2* & 2* & 2* & yes \\
\hline 
\end{tabular}
\end{center}
\caption{eqBT for triplets with arbitrary functions listed in Figure
  \ref{F:lim2}.}
\end{table}

\begin{table}
\begin{center}
\begin{tabular}{|l|c|c|c|c|c|c|c|}
\hline
Eqn &{\rotatebox[origin=c]{90}{{\scriptsize Top}}} & 
{\rotatebox[origin=c]{90}{{\scriptsize Bottom}}} &
{\rotatebox[origin=c]{90}{{\scriptsize Left}}} &
{\rotatebox[origin=c]{90}{{\scriptsize Right}}} &
{\rotatebox[origin=c]{90}{{\scriptsize  Back}}} &
{\rotatebox[origin=c]{90}{{\scriptsize  Front}}} & 
{\rotatebox[origin=c]{90}{{\scriptsize  param}}}\\
\hline
63,63,63-1 \eqref{636363-1}  & 1 & 1 & 1 & 1 & 1 & 1 & $p,q,r$\\
63,59,31-1 \eqref{635931-1} & $2_3$ & $2_1$ & 1 & 1 & 1 & 1 & $p,q$\\
\hline
63,63,63-2 \eqref{636363-2}   & 1 & 1 & 1 & 1 & 1 & 1 & $p,q,r$\\
63,59,31-2 \eqref{635931-2}  & 1 & 1 & 1 & 1 & 1 & 1 & $p,q$\\
63,58,15\phantom{-2}  \eqref{635815} & 2* & 2* & 1 & 1 & 1 & 1 & $p,q$\\
\hline
27,31,59\phantom{-2}  \eqref{273159} & 1 & 1 & 1 & 1 & 1 & 1 & $r$\\
26,15,59\phantom{-2}  \eqref{261559} & 1 & 1 & 2* & 2* & 1 & 1 & $r$\\
10,15,58\phantom{-2}  \eqref{101558} & 0 & 0 & 2* & 2* & 2* & 2* & $r$\\
\hline
53,53,53\phantom{-2}  \eqref{535353} & 1 & 1 & 1 & 1 & 1 & 1 & $p,q,r$\\
53,49,21\phantom{-2}  \eqref{534921}  & 1 & 1 & 1 & 1 & 1 & 1 & $p,q$\\
\hline
\end{tabular}
\end{center}
\caption{eqBT for triplets in Figure \ref{F:lim1}, with specific
  $p,q,r$ dependency listed. As mentioned they are all reductions of
  \{63,63,63-2\}. For \{63,59,31\}-1 the subscript gives the degree of
  the other equation. All of these equations have the tetrahedron
  property. Note that the reduction \{10,15,58\} can have free $c_i$
  functions and therefore it fails some eqBTs, which is indicated by
  the 0 entries.}
\end{table}

\begin{samepage}
Here are some examples
\begin{itemize}
\item For equation \{58,10,10\} the bottom-back-top-front side
  equations produce for the left equation a rational expression
  whose numerator factorizes as
\[
 y\,(\t x \b x - x\xbt x)\left[c_6(q,r)(c_2(p,q) \xbt x + c_5(p,q) \b x)\,\t x
 + c_5(p,q)(c_4(p,q) \b x + c_6(q,r) \xbt x)\,x\right]
\]
The factor in round brackets is the expected left equation. Note that
for the alternate equation in square brackets the $q$ dependence is
superfluous since the back equation should only depend on $r,p$.
\item For equation \{10,15,58\} the  the bottom-back-top-front side
  equations produce 
\[
 y\,(\t x \b x - x\xbt x)\left[\b x \t x+x \xbt x + r(x \t x + \b x
   \xbt x)\right]
\]
The term in square brackets is the proper left equation, the extra
factor $\t x \b x - x\xbt x$ is simpler and appears often. For this case
the same type of result is found for right, back and front
equations. All ``2*'' cases are like this.
\item For \{63,59,31\}-1 eqBT produces the bottom equation and as an
  alternative, $x \h x\xht x + x\h x\t x + x\t x\xht x + \h x \t x
  \xht x$, while for the top equation the alternative is $y+\t y+\h
  y+\xht y$.
\end{itemize}
\end{samepage}

\subsection*{Summary}
All the equations discussed here satisfy the
consistency-around-the-cube condition. The equations on the
consistency cube can also provide a Lax pair or produce a B\"acklund
transformation in which the ``side equations'' may generate the bottom
and top equations. It turns out that many of the equations produce a
fake Lax pair or equivalently an empty eqBT. From the tables we can
also read that some equations have the tetrahedron property and still
fail with respect to eqBT. Furthermore some equations have ambiguous
eqBT, generating two possible equations. If such a triplet does not
completely fail in any direction we may have an example of partial
integrability.

\section{Discussion}
We have searched for all homogeneous quadratic triplets of multilinear
irreducible equations that satisfy the Consistency-Around-a-Cube
condition with uniform embedding (translation invariance of quad
equations). The results were given in Appendix \ref{S:resu} and
classified in Section \ref{S:class}. The three equations forming the
triplet were allowed to have different forms, while in the usual setup
the bottom equation is given and the other equations are obtained
from that by definite variable and parameter changes. In that sense
our approach is close to that of Boll \cite{Boll2011,Boll2012}.

The ansatz for the equations contained coefficients that depended on
the two lattice parameters associated with the quadrilateral in
question. The results can be divided into two classes: those that
cannot have arbitrary dependence on the two lattice parameters and
those that can. The arbitrary functions can sometimes be eliminated by
a gauge transformation (that preserves uniformity) and therefore the
{\em possibility} of free functions is essential. The results can be
arranged into sub-cases, some relations are given in Figures
\ref{F:lim2}-\ref{F:lim1}.  The equations that cannot have free
functions are all reductions of $Q3(\delta=0)$ of the ABS list
\cite{ABS03}.

Recently it has been noted that some equations that pass the CAC test
can have fake Lax pairs \cite{ButHay13}, and independently the
possibility of ``weak Lax pairs'' has also been observed
\cite{HieViall12}. In order to characterize the equations further we
studied how the B\"acklund transformation works on these
equations. The result are given in Tables 1-3. One finds that all
equations that {\em can} have free functions also have some failing
BTs (or equivalently some fake Lax pairs). This is natural since it is
not possible to generate a free function from equations that do not
contain them. Note that some of the equations with failing eqBTs do
nevertheless have the tetrahedron property.

From our results we infer the following:
\begin{enumerate}
\item The CAC condition can only be a {\em necessary} condition of
  integrability. It cannot be sufficient as many equations with
  failing eqBTs pass this test.

\item The CAC condition may be sufficient if accompanied with some
  other type of condition, but the tetrahedron condition is not enough
  for that purpose.

\item We conjecture that if there is a unique BT for each direction of
  the cube then that should be sufficient for integrability.
  Non-uniqueness in some directions may be a signal of partial
  integrability while failure of the BT in one or more directions
  suggests non-integrability.

\item Integrability may be lost during reductions. But note that the
  reduction here does not change dimension but just means
  simplifying the equation by some limit of the coefficient(s).
\end{enumerate}

\appendix

\section{Results of the search\label{S:resu}}
We will now list the CAC equation triplets within the category of
equations studied in this paper, namely those that are quadratic and
acceptable by Definition \ref{D:1}. We list them following the order
of the search process which was based on the triplet code. In Section
\ref{S:class} they are grouped according to the ``highest equation'' and
its reductions.  During the search process sub-case inclusion was not
considered except for obvious reductions. If the triplet found is a
member of a symmetry orbit of more than one entry we also list the
orbit.

\subsection{{\bf\{10,10,10\}}\label{S:A.1}}
The triplet
\bse\label{101010-free}\begin{eqnarray}
c_2(p,q) \t x\h x + c_4(p,q) x \xht x&=&0,\\
c_2(q,r) \h x\b x + c_4(q,r) x \xbh x&=&0,\\
c_2(r,p) \b x\t x + c_4(r,p) x \xbt x&=&0,
\end{eqnarray}\ese
satisfies the CAC condition without any constraints on the parameters
$c_2,c_4$.  However, by the gauge transform \eqref{eq:fullgauge} we can
take the triplet into the simple form
\bse\label{101010}\begin{eqnarray}
\t x\h x - x \xht x&=&0,\\
\h x\b x - x \xbh x&=&0,\\
\b x\t x - x \xbt x&=&0,
\end{eqnarray}\ese
with
\[
\xbht x=\frac{\t x\h x\b x}{x^2},
\]
which does not have the tetrahedron property.

\subsection{$\{X,10,10\},\,X\neq 10 $\label{S:A.2}}
This category contains the triplet {\bf\{63,10,10\}} and its special
cases.  \bse\label{A631010}\begin{eqnarray} x\t x\,c_1(p,q)+\t x \h x
  c_2(p,q)+\h x\xht x\,c_3(p,q)+ x\xht x\, c_4(p,q)+x \h x c_5(p,q)+
  \t x\xht x c_6(p,q)&=&0,\phantom{mmm}
\label{A631010bot}\\
\h x \b x - x\xbh x&=&0,\label{A631010bac}\\
\b x \t x - x\xbt x&=&0,\label{A631010lef}
\end{eqnarray}
\ese 
That is, a completely arbitrary homogeneous quadratic equation is
MDC with the side equations as given above. The triply
shifted $x$ is given by
\[
\xbht x =-\frac{ c_5(p,q)x\b x\h x+c_1(p,q)x \t x \b x
  +c_2(p,q)\t x\h x\b x}{c_4(p,q)x^2+c_6(p,q)x\t x+c_3(p,q)x\h x}\,,
\]
and thus \eqref{A631010} has TET in the sub-case
$c_2(p,q)=c_4(p,q)=0$, with code \{53,10,10\}.

The sub-cases of \{63,10,10\} arrange themselves into orbits as follows:
\begin{center}
\begin{tabular}{l  l}
6 terms: & \{63,10,10\}\\
5 terms: & \{55,10,10\},\{61,10,10\},\\
5 terms: & \{31,10,10\},\{47,10,10\},\{59,10,10\},\{62,10,10\}\\
4 terms: & \{53,10,10\},\\
4 terms: & \{15,10,10\},\{58,10,10\},\\
4 terms: & \{27,10,10\},\{30,10,10\},\{43,10,10\},\{46,10,10\},\\
4 terms: &
\{23,10,10\},\{45,10,10\},\{29,10,10\},\{51,10,10\},\{57,10,10\},\\
& \{60,10,10\},\{39,10,10\},\{54,10,10\},\\
3 terms: & \{7,10,10\},\{13,10,10\},\{50,10,10\},\{56,10,10\},\\
3 terms: & \{11,10,10\},\{14,10,10\},\{26,10,10\},\{42,10,10\},\\
3 terms: & \{21,10,10\},\{37,10,10\},\{49,10,10\},\{52,10,10\},\\
2 terms: & \{5,10,10\},\{48,10,10\}.
\end{tabular}
\end{center}
In solving the CAC equations for these triplet codes we were usually
led to conditions on the $c_2,c_4$ terms of the back and left
equation.  For \{63,10,10\} it is not possible to have any free
functions in the side equations, but as the \{58,10,10\} \eqref{581010}
case shows, with smaller number of terms in the bottom equation we
have more freedom in the side equations but that freedom can be
eliminated by gauge. 

\subsection{$\{10,X,Y\},\,X,Y\neq 10 $\label{S:3.2.3}}
In this case we arrange the bottom equation to be the simple one.
The main result in this category is {\bf\{10,58,15\}}:
\bse\label{A105815}\begin{eqnarray} \h x\t x-x\xht x=0,\\
 x \b x
c_5(q,r) + x \xbh x c_4(q,r) + \b x \h x c_2(q,r) + \xbh x \h x
c_6(q,r)=0,\\ x \b x c_1(r,p) + x \xbt x
c_4(r,p) + \b x \t x c_2(r,p) + \xbt x \t x c_3(r,p)=0.
\end{eqnarray}
\ese The triply shifted quantity is given by
\begin{equation}\label{eq:trp-105815}
\xbht x=\b x\,\frac{(c_1(r,p)x+c_2(r,p)\t x)(c_5(q,r)x+c_2(q,r)\h x)}
{(c_4(r,p)x+c_3(r,p)\t x)(c_4(q,r)x+c_6(q,r)\h x)}.
\end{equation}
This has sub-cases when one or two $c_i$ in either or both equations
vanish while keeping the equations irreducible. Missing $c_j$ terms
sometimes allow a more general bottom equation, but that freedom can be
gauged away. The sub-case orbits are as follows:

\begin{center}
\begin{tabular}{l  l}
2,4,4 terms: & \{10,58,15\},\\
2,3,4 terms: & \{10,58,7\},\{10,58,13\},\{10,50,15\},\{10,56,15\},\\
2,3,4 terms: & \{10,58,11\},\{10,58,14\},\{10,26,15\},\{10,42,15\},\\
2,3,3 terms: & \{10,26,11\},\{10,26,14\},\{10,42,11\},\{10,42,14\},\\
2,3,3 terms: & \{10,50,11\},\{10,50,14\},\{10,56,11\},\{10,26,7\},\{10,26,13\},
\{10,42,7\},\\&
 \{10,56,14\},\{10,42,13\},\\
2,3,3 terms: & \{10,50,7\},\{10,50,13\},\{10,56,13\},\{10,56,7\},\\
2,2,4 terms: & \{10,58,5\},\{10,48,15\},\\
2,2,3 terms: & \{10,56,5\},\{10,50,5\},\{10,48,13\},\{10,48,7\},\\
2,2,3 terms: & \{10,26,5\},\{10,42,5\},\{10,48,11\},\{10,48,14\},\\
2,2,2 terms: & \{10,48,5\}.
\end{tabular}
\end{center}
The number of terms in each equation is given in the first column,
but that list is not ordered because some reflections exchange back
and left equations. The sub-cases belonging to \{X,10,10\} are
excluded.  

\tdplotsetmaincoords{75}{115}
\begin{figure}
\centering
\begin{tikzpicture}[scale=1.50,tdplot_main_coords, 
cube/.style={very thick,black}]  
\crosspq
\crossqr
\crossrp
\edgespr{r}{0}
\edgesqr{r}{0}
\laatikko
\node at (0.5,0.5,-1) {\eqref{A105815}};
\end{tikzpicture}
\hspace{1cm}\begin{tikzpicture}[scale=1.50,tdplot_main_coords, 
cube/.style={very thick,black}]
\crosspq
\crossqr
\crossrp
\edgesqr{q}{0}
\edgespr{p}{0}
\laatikko 
\node at (0.5,0.5,-1) {\eqref{101558}};
\end{tikzpicture}
\caption{A diagrammatic way of describing the nonzero terms in the
  triplets \eqref{A105815} and \eqref{101558}: each line or band
  connecting two points corresponds to their product term in the
  equation.\label{F:1}}
\end{figure}
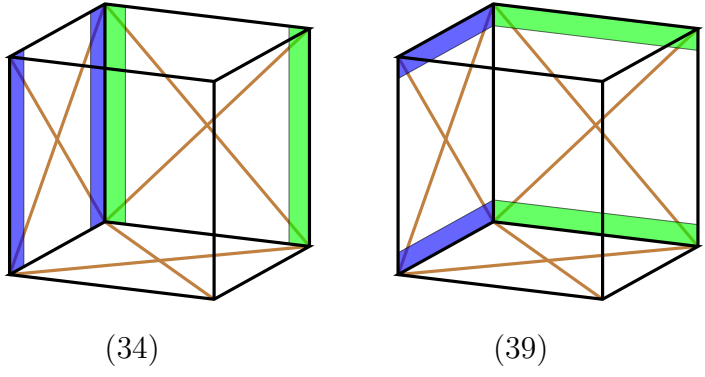

\paragraph{\{10,15,58\}} 
In this category there is also an equation that is not a sub-case of
\eqref{A105815}:
\bse\label{A101558}\begin{eqnarray}
\h x\t x+x\xht x c_4(p,q)=0,\\
 \b x \h x + x \xbh x + x \h x g(r)^{-1} + \b x \xbh x f(r)^{-1}=0.\\
 \b x \t x + x \xbt x +x \t x f(r) + \b x \xbt x g(r)=0,
\end{eqnarray}
\ese
with 
\begin{equation}\label{101558triple}
\xbht x=-\frac{\t x\h x f(r)}{\b x c_4(p,q) g(r)}
\end{equation}
This cannot be a sub-case of \eqref{105815} because the equations do
not contain the same shifted $x$ variables.  We can use gauge
transformation to simplify this further. Choosing the gauge so that
\eqref{101558triple} simplifies to
\begin{equation}\label{101558triple2}
\xbht x=-\frac{\t x\h x}{\b x}
\end{equation}
the equations get the form (after redefining $r$)
\bse\label{101558}\begin{eqnarray} 
\h x\t x-x\xht x=0,\\ 
 r(\b x \h x+x \xbh x) + x \h x + \b x \xbh x=0,\\
 \b x \t x+x \xbt x + r(x \t x + \b x \xbt x) =0.
\end{eqnarray}
\ese

For an intuitive understanding of the difference between
\eqref{A105815} and \eqref{101558} we can look at the Figure
\ref{F:1}. Note also that in \eqref{eq:trp-105815} $\b x$ is in the
numerator while in \eqref{101558triple2} it is in the denominator. 
 
\subsection{$\{58,15,X\},\,X\neq 10 $}
Only {\bf\{63,58,15\}} survives, all other cases lead to equations
that factorize.  \bse\label{635815}\begin{eqnarray} p (1 - q^2)(x \h x
  + \t x \xht x) - q (1-p^2)(x \t x + \h x \xht x) + (p^2 - q^2)( x
  \xht x + \h x \t x )&=&0,\\ q (x \b x + \h x \xbh x) + x \xbh x + \b
  x \h x&=&0,\\ p (x \b x + \t x \xbt x) + x \xbt x + \b x \t x&=&0.
\end{eqnarray}\ese 
\[
\xbht x=\frac{\b x ( - p \h x + q \t x)}{p \t x - q \h x}.
\]

\subsection{$\{15,58,X\},\,X\neq 10,15,58 $\label{S:1558}}
According to \eqref{eq:thdef1} we have by tilde-hat reflection and
rotation $\{15,58,X\} \to \{58,\theta X,15\} \to \{15,58,\theta
X\}$. Thus $\{15,58,X\}$ and $ \{58,15,Y\}$, are not related by this
reflection and may have different types of solutions.
 
The main solution in this category is {\bf{\{53,15,58\}}}, and after a
gauge transformation the equation satisfying CAC can be written as
\bse\label{A531558}\begin{eqnarray} c_6(p,q) \xht x \t x + c_3(p,q) \h
  x \xht x + c_1(p,q) x \t x + c_5(p,q) x \h x&=&0,\\ r(x \h x + \b x
  \xbh x) + x \xbh x + \b x \h x&=&0,\\ r(x \t x + \b x \xbt x) + x
  \xbt x + \b x \t x&=&0.
\end{eqnarray}\ese
The triply shifted variable is
\[
\xbht x=-\b x\,\frac{c_1(p,q) \t x + c_5(p,q) \h x }
{c_6(p,q) \t x + c_3(p,q) \h x}\,,
\]
thus it has TET.  All other equations in this
category that satisfy CAC are sub-cases of the above, or lead to
equations that factorize. The orbits for the sub-cases are
 \begin{center}
\begin{tabular}{l  l}
4 terms: & \{53,15,58\},\\
3 terms: & \{21,15,58\},\{37,15,58\},\{49,15,58\},\{52,15,58\},\\
2 terms: & \{5,15,58\},\{48,15,58\}.
\end{tabular}
\end{center}

\subsection{$\{15,X,Y\},\,X,Y\not\in \{10,58\}$ or $\{58,X,Y\},
 \,X,Y\not\in \{10,15\}$}
By the tilde-hat reflection we have $\{15,X,Y\}\to \{58,\theta
Y,\theta X\}$ and therefore there is one-to-one correspondence between
the sets $\{15,X,Y\}$ and $\{58,X',Y'\}$ and we only need to discuss
one of them.

There are types two solutions, the first one is {\bf \{58,48,5\}} given by
\bse\label{584805}\begin{eqnarray}
   c_4(p,q)(\sigma_1 \t x \h x + x \xht x)
 + c_5(p,q)(\sigma_2  x \h x+ \t x\xht x)&=&0,\\
c_5(q,r) x\b x + c_6(q,r) \h x \xbh x&=&0,\\
x\b x-\sigma_1\sigma_2\t x \xbt x&=&0,
\end{eqnarray}\ese
with
\[
\xbht x=\frac{x\b x\,c_5(q,r)}{\h x\,c_6(q,r)}
\frac{c_4(p,q) x + c_5(p,q)\t x}
{c_4(p,q)\sigma_2 \t x  + \sigma_1c_6(p,q) x}.
\]
One can change some of the signs by gauge but one cannot eliminate the
freedom in $c_i$.  This does not have TET except if $c_4(p,q)=0$, in
which case it becomes \{48,48,5\} \eqref{484805}. The reflected case
is \{15,48,5\}.

\paragraph{\{26,15,59\}} The second solution is, after gauge and rotation
\bse\label{261559}\begin{eqnarray} 
x\h x+x\xht x+\t x\h x&=&0,\\
r(x \h x+\xbh x\b x)+x\xbh x+\b x\h x&=&0,\\
(r^2-1)x\b x + r^2(x\xbt x+\b x\t x)+r(x\t x+\xbt x\b x)&=&0,
\end{eqnarray}\ese 
\[
\xbht x=\frac{\h x(\b x-r\t x)}{r\b x}
\] 
The triplet \eqref{261559} is symmetric under bar reversal and tilde
reversal, but not under hat-reversal and therefore there are also the
corresponding solution within three reversed categories. The orbit is
\{15,59,26\},\{15,62,42\},\{58,11,31\},\{58,14,47\}.

\subsection{$\{53,X,Y\},\,X,Y\not\in \{10,15,58\}$}
There are five solutions in this category and they all have TET.

\paragraph{ \{53,48,5\}}
\bse\label{534805}\begin{eqnarray} 
c_5(p,q) (x \h x -\sigma_1\sigma_2 \xht x \t x)
 + c_1(p,q) ( x \t x  -\sigma_1\sigma_3  \h x \xht x)&=&0,\\
x \b x - \sigma_3 \xbh x \h x&=&0,\\
x \b x - \sigma_2 \xbt x \t x&=&0,
\end{eqnarray}\ese
where $\sigma_j^2=1$. The signs $\sigma_j$ can be changed by gauge.
\[
\xbht x=\frac{\b x (c_1(p,q) \sigma_2 \h x + c_5(p,q) \sigma_3 \t
  x)} {\sigma_1 (c_1(p,q) \t x + c_5(p,q) \h
  x)}.
\]

\paragraph{\{53,5,48\}}
\bse\label{530548}\begin{eqnarray} 
 c_1(p,q) x \t x + c_3(p,q)\h x\xht x + c_5(p,q) x \h x +
  c_6(p,q) \t x \xht x)&=&0,\\
x \h x  + \xbh x \b x&=&0,\\
x \t x + \xbt x \b x&=&0,
\end{eqnarray}\ese
\[
\xbht x=-\b x\,\frac{c_1(p,q)\t x+c_5(p,q)\h x}
{c_6(p,q)\t x+c_3(p,q)\h x}.
\]
The sub-case orbits are as for \{53,15,58\} \eqref{A531558}.

\begin{figure}
\centering
\begin{tikzpicture}[scale=1.50,tdplot_main_coords, 
cube/.style={very thick,black}]  
\edgespr{r}{0}
\edgesqr{r}{0}
\edgespq{q}{0}
\edgespq{r}{0}
\laatikko
\node at (0.5,0.5,-1) {\eqref{534805}};
\end{tikzpicture}
\hspace{1cm}\begin{tikzpicture}[scale=1.50,tdplot_main_coords, 
cube/.style={very thick,black}]
\edgesqr{q}{0}
\edgespr{p}{0}
\edgespq{q}{0}
\edgespq{r}{0}\laatikko 
\node at (0.5,0.5,-1) {\eqref{530548}};
\end{tikzpicture}
\caption{Diagrams for the nonzero terms in the
  triplets \eqref{534805} and \eqref{530548}.
\label{F:58}}
\end{figure}
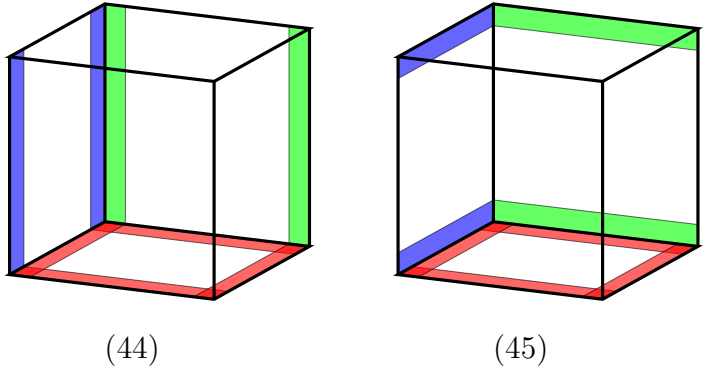

The difference between \eqref{534805} and \eqref{530548} is
illustrated in Figure \ref{F:58}.

\tdplotsetmaincoords{75}{115}

\paragraph{ \{53,53,53\}}
\bse\label{535353}\begin{eqnarray} 
p (x \h x \delta_p+\t x \xht x \rho_p)
-q (x \t x \delta_q+\h x \xht x \rho_q)&=&0,\\
q (x \b x \delta_q+\h x \xbh x \rho_q)
-r (x \h x \delta_r+\b x \xbh x \rho_r)&=&0,\\
r (x \t x \delta_r+\b x \xbt x \rho_r)
-p (x \b x \delta_p+\t x \xbt x \rho_p)&=&0,
\end{eqnarray}\ese
with triply shifted variable
\[
\xbht x=\frac{\b x \h x \delta_p p (\rho_q \delta_q q^2 - \rho_r
  \delta_r r^2) + \b x \t x \delta_q q ( \rho_r \delta_r r^2 - \rho_p
  \delta_p p^2) + \h x \t x \delta_r r (\rho_p \delta_p p^2 - \rho_q
  \delta_q q ^2)}{\t x \rho_p p (\rho_r \delta_r r^2 - \rho_q \delta_q
  q^2)+ \h x \rho_q q (\rho_p \delta_p p^2 - \rho_r \delta_r r^2)+\b x
  \rho_r r (\rho_q \delta_q q^2 - \rho_p \delta_p p^2)}.
\]
 We have added here some coefficients $\delta_\alpha$ and
 $\rho_\alpha$ and if they are nonzero they can be gauged to $1$,
 which yields $H3(\delta=0)$ of the ABS list. If a $\rho_\alpha$
 vanishes we get a rotation of \{53,49,21\}, if a $\delta_\alpha$
 vanishes we get a rotation of its reflection \{53,52,37\}, and if
 $\delta_\alpha=\rho_\alpha=0$ for some $\alpha$ we get a rotation of
 a special case of \{53,48,05\}.

\paragraph{\{53,49,21\}}
\bse\label{534921}\begin{eqnarray} q(x \t x + \h x\xht x) -p (x \h x +
  \t x\xht x)&=&0,\\ x \h x - q(x \b x + \h x\xbh x)&=&0,\\ x \t x -
  p(x \b x + \t x\xbt x)&=&0,
\end{eqnarray}\ese
\[
\xbht x=\frac{pq\b x(q\h x-p\t x)+(p^2-q^2)\h x\t x}{pq(p\h x-q\t x)}.
\]
There is also the reflected case \{53,52,37\}.

\paragraph{\{53,11,26\}}
\bse\label{531126}\begin{eqnarray} 
 c_1(p,q) x \t x + c_3(p,q)\h x\xht x + c_5(p,q) x \h x +
  c_6(p,q) \t x \xht x)&=&0,\\
x \h x + \b x\h x + \xbh x x&=&0,\\
x \t x + \b x\t x + \xbt x x&=&0,
\end{eqnarray}\ese
\[
\xbht x=-\b x\,\frac{c_1(p,q)\t x+c_5(p,q)\h x}
{c_6(p,q)\t x+c_3(p,q)\h x}.
\]
Note that $\xbht x$ is the same as for \{53,5,48\} \eqref{530548}.
There is also the reflected case \{53,14,42\}. Sub-case orbits are as
for \{53,15,58\} \eqref{531558}.

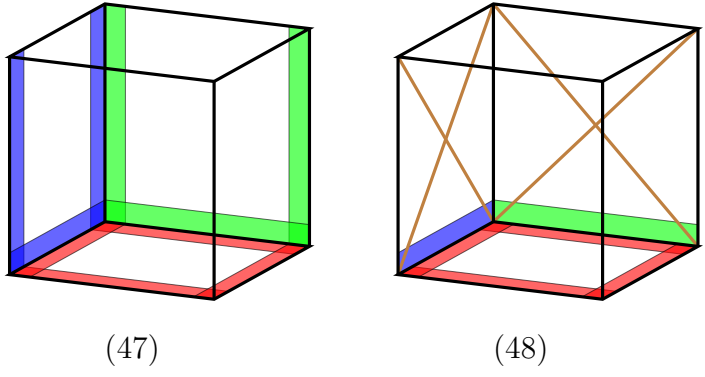
\begin{figure}
\centering
\begin{tikzpicture}[scale=1.50,tdplot_main_coords, 
cube/.style={very thick,black}]  
\edgespr{r}{0}
\edgesqr{r}{0}
\edgespq{q}{0}
\edgespq{r}{0}
\yzband{0}{0}{0}{2}{0.2}{green}{0.6};
\xzband{0}{0}{0}{2}{0.2}{blue}{0.6};
\laatikko
\node at (0.5,0.5,-1) {\eqref{534921}};
\end{tikzpicture}
\hspace{1cm}\begin{tikzpicture}[scale=1.50,tdplot_main_coords, 
cube/.style={very thick,black}]
\yzband{0}{0}{0}{2}{0.2}{green}{0.6};
\xzband{0}{0}{0}{2}{0.2}{blue}{0.6};
\crossqr
\crossrp
\edgespq{q}{0}
\edgespq{r}{0}\laatikko 
\node at (0.5,0.5,-1) {\eqref{531126}};
\end{tikzpicture}
\caption{Diagrams for the nonzero terms in the
  triplets \eqref{534921} and \eqref{531126}.
\label{F:53}}
\end{figure}

The difference between \eqref{534921} and \eqref{531126} is
illustrated in Figure \ref{F:53}.

\tdplotsetmaincoords{75}{115}

\subsection{$\{5,X,Y\}$ or $\{48,X,Y\}\,X,Y\not\in
 \{10,15,53,58\}$\label{S:A.8}}
By the tilde-hat reflection we have $\{5,X,Y\}\to \{48,\theta Y,\theta
X\}$.  The solutions in this category are as follows:

\paragraph{ \{48,48,05\}}
\bse\label{484805}\begin{eqnarray} x \h x - \sigma_1 \t x \xht x
  &=&0,\label{484805-2}\\ c_5(p,q) x \b x+c_6(p,q) \h x\xbh x&=&0,\\ x
  \b x - \sigma_2\t x \xbt x &=&0,\label{5548-3}
\end{eqnarray}\ese
\[ 
\xbht x=-\sigma_1\sigma_2\t x\, \frac{\b x c_5(q,r)}{\h x c_6(p,q)}.
\]
The signs can be controlled by gauge. In this orbit there is also
\{05,05,48\}.

\paragraph{\{21,05,48\}}
\bse\label{210548}\begin{eqnarray} 
c_1(p,q) x \t x+c_3(p,q) \h x\xht x+c_5(p,q)x \h x&=&0,\\
x \h x + \b x \xbh x &=&0,\label{210548-3}\\
x \t x + \b x \xbt x &=&0\label{210548-2},
\end{eqnarray}\ese
\[ 
\xbht x=-\b x\frac{ \t x c_1(q,r)+\h x c_5(p,q) }{\h x c_3(p,q)}.
\]
The full orbit of reflected cases is
\{21,5,48\}, \{37,5,48\}, \{49,5,48\}, \{52,5,48\}.

\paragraph{{\{63,48,5\}}} \bse\label{A634805}\begin{eqnarray}
 c_1(p,q)(x \t x -  \sigma_1 \sigma_3 \h x \xht x)
 + c_5(p,q) ( x \h x - \sigma_1 \sigma_2 \xht x \t x)
 + c_4(p,q) (\h x \t x - \sigma_1 x \xht x)&=&0,\\
x \b x - \sigma_3 \xbh x \h x&=&0,\\
x \b x - \sigma_2 \xbt x \t x&=&0,
\end{eqnarray}\ese
\[
\xbht x=\frac{x \b x (c_1(p,q) \sigma_3 \h x + c_4(p,q) x + c_5(p,q)
  \sigma_2 \t x)} {\sigma_1 \sigma_2 \sigma_3 (c_1(p,q) x \t x +
  c_4(p,q) \h x \t x + c_5(p,q) x \h x)}.
\]
The sign $\sigma_1$ cannot be eliminated by gauge.

\paragraph{{\{5,11,26\}}}
\bse\label{051126}\begin{eqnarray} 
c_1(p,q) x \t x+c_3(p,q) \h x\xht x&=&0,\\
x \h x + \b x \h x+ x \xbh x &=&0,\\
x \t x + \t x \b x + x \xbt x &=&0,
\end{eqnarray}\ese
\[
\xbht x=-\frac{\t x \b x c_1(q,r)}{\h x c_3(p,q)}.
\]
The orbit is \{5,11,26\}, \{48,11,26\}, \{5,14,42\}, \{48,14,42\}.

\subsection{The rest:$\{X,Y,Z\},\, X,Y,Z \not\in\{5,10,15,53,58\}$}
\paragraph{ \{27,31,59\}}
\bse\label{273159}\begin{eqnarray}
x \h x + x \t x + x \xht x + \h x \t x&=&0,\\
 (1-r^2)\, x \b x + (x \xbh x + \b x \h x) +r(x \h x+\b x \xbh x)&=&0,\\
 (r^2-1)\, x \b x+r^2(x
\xbt x + \b x \t x)+r (x \t x+\b x \xbt x)&=&0,
\end{eqnarray}\ese
\[
\xbht x=\frac{r^2\b x \t x - r \h x \t x + \b x \h x}{r \b x}.
\]
There are altogether 4 different reflections of this triplet:
\{27,31,59\}, \{43,31,62\}, \{30,47,59\}, \{46,47,62\}.

\paragraph{ \{21,11,26\}}
\bse\label{211126}\begin{eqnarray}
 x \h x c_5(p,q) + x \t x c_1(p,q) + \h x  \xht x c_3(p,q)&=&0,\\
x \h x + x \xbh x + \b x \h x&=&0,\\
x \t x + x \xbt x + \b x \t x&=&0.
\end{eqnarray}\ese
\[
\xbht x=-\frac{\t x \b x c_1(q,r)+\h x\b x c_5(p,q)}{\h x c_3(p,q)}.
\]
There are 8 elements in the orbit \{11,26,21\}, \{11,26,37\},
\{11,26,49\}, \{11,26,52\}, \{14,42,21\}, \{14,42,37\}, \{14,42,49\},
\{14,42,52\};

\vskip 0.5cm The remaining cases were the difficult ones to solve. In the
category \{63,59,31\} there are two solutions:

\paragraph{ \{63,59,31\}-1}

\bse\label{635931-1}\begin{eqnarray} (q - p) (x \xht x + \h x \t x) -
  p (x \t x + \h x \xht x) + q (x \h x + \xht x \t x)&=&0,\\ q x \h x
  + (x \b x + \xbh x \h x) + (x \xbh x + \b x \h x) &=&0,\\ p x \t x +
  (x \b x + \xbt x \t x) + (x \xbt x + \t x \b x) &=&0,
\end{eqnarray}\ese
\[
\xbht x=\b x+\frac{(p-q) \h x \t x}{\h x-\t x}.
\]
Note that the back and left equations are not related by a cyclic
variable change.

\paragraph{ \{63,59,31\}-2}

\bse\label{635931-2}\begin{eqnarray}
(p^2-q^2) (x \xht x + \h x \t x) + (p^2-1) q (x \t x + \h x \xht x) -
 p (q^2-1) (x \h x + \xht x \t x) &=&0,\\
(q^2-1) x \h x + q (x \b x + \xbh x \h x) + x \xbh x + \b x \h x&=&0,\\
(p^2-1) x \t x + p (x \b x + \xbt x \t x) + x \xbt x + \b x \t x&=&0,
\end{eqnarray}\ese
\[
\xbht x=\frac{(q^2-p^2) \h x \t x+\b x (q \t x-p \h x)}{p \t x-q \h x}.
\]
There are also corresponding two reflected cases \{63,62,47\}.

\paragraph{ \{63,63,63\}}
In the category \{63,63,63\} there are also two solutions. The
preliminary way of writing the result, after a gauge transformation,
is the form
 \bse\label{636363-alku}\begin{eqnarray} c(q) (x \t x + \h
  x \xht x) - c(p) (x \h x + \xht x \t x) - p (x \xht x + \h x \t x) +
  q (x \xht x + \h x \t x)&=&0,\\ c(r) (x \h x + \b x \xbh x) - c(q)
  (x \b x + \xbh x \h x) - q (x \xbh x + \b x \h x) + r (x \xbh x + \b
  x \h x)&=&0,\\ c(r) (x \t x + \b x \xbt x) - c(p) (x \b x + \xbt x
  \t x) - p (x \xbt x + \b x \t x) + r (x \xbt x + \b x \t x)&=&0.
\end{eqnarray}\ese

where we still have the conditions
\[
c(\alpha)^2=\alpha^2+a_1 \alpha+a_0,\quad \alpha\in\{p,q,r\}.
\]
If $a_1$ and/or $a_0$ are nonzero the result is awkward, but we can
simplify the result by a transformation $p=f(P),\,q=f(Q),\,
r=f(R)$. However, in order to connect with known results we we want
the coefficient of the $(x \xht x + \h x \t x)$ be $(P^2-Q^2)$ after
the transformation and a possible overall multiplication. For
simplicity let us redefine $a_0=a_1^2/4-a_2^2$. There are two
possibilities:

\paragraph{ \{63,63,63\}-1} If $a_2=0$ then by a simple translation 
we get $c(\alpha)=\epsilon(\alpha) \alpha$, where
$\epsilon(\alpha)^2=1$, but this sign can be eliminated by gauge. In
order to connect with known results we redefine $p\mapsto
1/p,\,q\mapsto 1/q,\,r\mapsto 1/r$ after which we get
\bse\label{636363-1}\begin{eqnarray} (q-p) (x \xht x + \h x \t x) - q
  (x \h x + \xht x \t x) + p (x \t x + \h x \xht x)&=&0,\\ (r-q) (x
  \xbh x + \b x \h x) - r (x \b x + \xbh x \h x) + q (x \h x + \b x
  \xbh x)&=&0,\\ (p-r) (x \xbt x + \t x \b x) - p (x \t x + \xbt x \b
  x) + r (x \b x + \t x \xbt x)&=&0,
\end{eqnarray}\ese
\[
\xbht x=\frac{r \b x (\h x - \t x) + q \h x (\t x - \b x) + p
  \t x (\b x - \h x)}{r (\h x - \t x) + q (\t x - \b x) + p
  (\b x - \h x)}.
\]
This is in fact Q1($\delta=0$) in the ABS list.

\paragraph{ \{63,63,63\}-2} A different solution is obtained if 
$a_2 \neq 0$. Then by transforming
 \[
p=\frac{( - a_1 + 2 a_2)P^2  + a_1 + 2 a_2}{2 P^2 - 2},
\] etc. one finds
\[
 c(P)=\epsilon_Pa_2\frac{ 2 P}{P^2-1},\text{ and }(p-q) \to 
 -2 a_2 \frac{P^2-Q^2}{(P^2-1)(Q^2-1)},
\] 
and after clearing denominators (and renaming $P\to p$) we get
$Q3(\delta=0)$ of the ABS list:
\bse\label{A636363-2}\begin{eqnarray}
(q^2 - p^2) (x \xht x + \h x \t x) - p (q^2-1) (x \h x + \xht x \t x)
+(p^2-1) q (x \t x + \h x \xht x)&=&0,\\ (r^2 - q^2) (x \xbh x + \b x
\h x) - q (r^2-1) (x \b x + \xbh x \h x) +(q^2-1) r (x \h x + \b x
\xbh x)&=&0,\\ (p^2 - r^2) (x \xbt x + \t x \b x) - r (p^2-1) (x \t x
+ \xbt x \b x) +(r^2-1) p (x \b x + \t x \xbt x)&=&0,
\end{eqnarray}\ese
with
\[
\xbht x=\frac{\b x \h x p (q^2 - r^2) + \b x \t x q (r^2 - p^2) +
 \h x \t x r (p^2 - q^2)}{\b x  r (q^2 - p^2) + \h x q (p^2 - r^2) +
 \t x p (r^2 - q^2)}\,.
\]

Note that the difference between Q1 and Q3 arises from the
factorization properties of the $c(\alpha)$ coefficient in
\eqref{636363-alku}.

\section*{Acknowledgment} I would like to thank Da-jun Zhang for 
useful comments on the manuscript. All computations were done using the
REDUCE computer algebra system \cite{REDUCE}

\end{document}